\documentclass[preprint,12pt]{elsarticle}

\usepackage{amssymb}
\usepackage{amsmath}
\usepackage[linesnumbered]{algorithm2e}

\usepackage{xcolor}

\newcommand{\blue}{}

\journal{Journal of Computational Physics}

\begin{document}

\begin{frontmatter}



\title{Moment-preserving particle merging via non-negative least squares} 


\author{Georgii Oblapenko, Manuel Torrilhon} 

\affiliation{organization={Applied and Computational Mathematics, RWTH Aachen},
            addressline={Schinkelstrasse 2}, 
            city={Aachen},
            postcode={520262}, 
            country={Germany}}

\begin{abstract}
A novel particle merging algorithm for rarefied gas dynamics simulations is proposed
that can conserve arbitrary velocity and spatial moments of the particle distribution
via solving a non-negative least squares problem. An extension that preserves
both exact and approximate collision rates
is also derived. The algorithm is applied to the simulation of several model
rarefied gas dynamics problems, where it exhibits noticeably lower merging-induced error
in key macroscopic quantities.
\end{abstract}




\end{frontmatter}



\section{Introduction}
\label{sec:intro}
Stochastic particle methods such as the Direct Simulation Monte Carlo (DSMC)~\cite{DSMC_Bird} and Particle-in-Cell~\cite{hockney1988computer}
are the method of choice for simulation of non-equilibrium rarefied gas and plasmas, where more simple models such as those
based on Euler and Navier--Stokes equations~\cite{colonna2022plasma} or moment methods~\cite{torrilhon2016modeling} are not applicable.
In such particle methods, the gas flow is modelled by a large number of computational particles, each of which represents
many actual physical particles such as atoms, molecules, or electrons. The number of physical particles represented by such a
computational particle is oftentimes referred to as the ``weight'' or ``computational weight'' of the particles.
These computational particles interact with each other through stochastic collisions
and/or self-induced electric and magnetic fields (in the case of Particle-in-Cell simulations).

However, in certain scenarios, the number of particles in a simulation needs to be reduced. This can be due to large density gradients in the system~\cite{massaro2024efficient} or occur in axisymmetric simulations with particles being weighted relative to the size of the computational cell to improve statistics in cells close to the rotational axis~\cite{hara2023effects}.
Additionally, in case particle-particle collisions are simulated, and the particles have non-equal computational weights,
these collisions lead to creation of new particles~\cite{rjasanow1996stochastic,schmidt2000new}, a process that leads to an exponential
growth in the number of particles if left unchecked. Additionally, some recent collision algorithms~\cite{araki2020interspecies,oblapenko2022hedging} rely on additional splitting of particles
in order to improve resolution of certain collisional processes, further accelerating particle number growth rates.
To remedy this, particles need to be removed from the simulation~\cite{lapenta1994dynamic,rjasanow1998reduction} in a process alternately referred to in the literature as
``particle merging'', ``particle coalescence'', ``particle reduction'', or ``particle number control''.
Such methods select a certain subset of particles to operate on, delete some of the particles and adjust the properties (computational weights, positions, velocities)
of the others. Since information is inevitably lost during such a procedure, the adjustment of the remaining particles' properties is done in order to restore physical consistency in the simulation,
such as ensuring conservation of mass, momentum, energy. Ensuring the conservation of only these properties is however insufficient, and merging can lead to noticeable errors in
simulations~\cite{oblapenko2022improving,hong2024improved}. Therefore, merging approaches aim to reduce the error they induce in the particle discretization, often done in a
semi-empirical manner by performing reduction on sub-groups of particles found via various clustering/binning approaches~\cite{vranic2015particle,frederiksen2015particle,martin2016octree,luu2016voronoi}.

Recent work has focussed on developing merging algorithms that explictly conserve higher-order properties of the system, such as the stress tensor and heat flux~\cite{lama2020higher,faghihi2020moment,goeckner2025generalized},
as well as higher-order moments~\cite{gonoskov2022agnostic,oblapenko2024non}, leading to improved simulation accuracy.
In the present work, we generalize the merging approach recently proposed in~\cite{oblapenko2024non}
to spatially inhomogeneous problems, improve the numerical stability of the algorithm by an appropriate scaling of the system, and derive
extensions that preserve both exact and approximate reaction rates. We compare this merging method with other approaches for several model
spatially homogeneous and one-dimensional flows of single-species gases and plasmas.

\section{Particle-merging}
\label{sec:particlemerging}
Particle merging can be defined in a general manner as an operator
\begin{equation}
  \mathcal{M}: \mathcal{W} \times \mathcal{V} \times \mathcal{X} \to \mathcal{W}' \times \mathcal{V}' \times \mathcal{X}',
\end{equation}
where $\mathcal{W} = \mathbb{R}_+^{N}$ and $\mathcal{W}' = \mathbb{R}_+^{M}$ are the pre- and post-merge weights of the particles correspondingly,  $\mathcal{V} = \mathbb{R}^{N \times d_v}$ and $\mathcal{V}' = \mathbb{R}^{M \times d_v}$ are the pre- and post-merge particles' velocities, and $\mathcal{X} = \mathbb{R}^{N \times d_x}$, $\mathcal{X}' = \mathbb{R}^{M \times d_x}$ are the pre- and post-merge particles' positions. Naturally, the post-merge number of particles is lower than the pre-merge one, i.e. $M<N$, and we will refer to the reduction procedure as ``$N:M$'' merging. In the present work we assume positive pre- and post-merge weights, although the extension of particle merging to negative weight particles, such as those describing deviations from equilibrium distributions rather than distributions directly, is an compelling topic for future research~\cite{parker1993fully}.
We consider the full 3-dimensional particle velocity vectors, that is, we take $d_v=3$, and in the present work we investigate spatially homogeneous ($d_x=0$) and one-dimensional ($d_x=1$) problems.

In this rather generic formulation, little can be said about the properties of $\mathcal{M}$ --- it oftentimes involves stochastic procedure to determine the post-merge properties, and it need not even be continuous in $\mathcal{W}$, $\mathcal{V}$, or $\mathcal{X}$. This is the case in binning-based approaches, where particles are grouped into bins according to some criteria, and a local merging procedure is applied in each bin. That is, a more simple local bin merging operator for a bin $b$ is defined:
\begin{equation}
  \mathcal{M}_b: \mathcal{W} \times \mathcal{V} \times \mathcal{X} \to \mathcal{W}_b' \times \mathcal{V}_b' \times \mathcal{X}_b',
\end{equation}
where $\mathcal{W}_b' = \mathbb{R}_+^{m_b}$,  $\mathcal{V}_b' = \mathbb{R}^{m_b \times d_v}$, $\mathcal{X}_b' = \mathbb{R}^{m_b \times d_x}$, and the post-merge number of bin particles $m_b$ is taken to be small so as to allow for analytical expressions of the post-merge particles' weights, velocities, and positions. Possible values of $m_b$ include 2, allowing for conservation of 1-st and 2-nd order non-mixed velocity and spatial moments~\cite{rjasanow1998reduction,martin2016octree}, and 6, allowing for conservation of the full stress tensor and heat flux~\cite{lama2020higher}.
Common binning approaches involve either a fixed grid in velocity space~\cite{vranic2015particle}, or adaptive octree structures~\cite{martin2016octree}; within a single grid cell spatial binning is usually not carried out, and only velocity information is used to group particles.

For example, in the case of an $N:2$ reduction procedure, one can compute the post-merge particles' (denoted with sub-indices 1 and 2) properties as
\begin{equation}
  w_1 = w_2 = \frac{1}{2} w_{\sum},
\end{equation}
\begin{equation}
  \mathbf{v}_1 = \mathbf{v}_{\sum} + \mathbf{u}_{v,3} \odot \mathbf{v}_{\text{std}}, \quad \mathbf{v}_2 = \mathbf{v}_{\sum} - \mathbf{u}_{v,3} \odot \mathbf{v}_{\text{std}},
\end{equation}
\begin{equation}
  \mathbf{x}_1 = \mathbf{x}_{\sum} + \mathbf{u}_{x,3} \odot \mathbf{x}_{\text{std}}, \quad \mathbf{x}_2 = \mathbf{x}_{\sum} - \mathbf{u}_{x,3} \odot \mathbf{x}_{\text{std}}.
\end{equation}
Here $\mathbf{u}_{v,3}$ and $\mathbf{u}_{x,3}$ are 3-dimensional vectors with random components of either -1 or 1, sampled randomly for each bin, and $\odot$ denotes the component-wise product.
The quantities with the $\sum$ indices denote average bin quantities, obtained by summation of a particle-specific property over all the pre-merge particles in a bin $b$:
\begin{equation}
  w_{\sum} = \sum_{i \in b} w_i,
\end{equation}
\begin{equation}
  \mathbf{v}_{\sum} = \frac{1}{w_{\sum}} \sum_{i \in b} w_i \mathbf{v}_i, \quad \mathbf{x}_{\sum} = \frac{1}{w_{\sum}} \sum_{i \in b} w_i \mathbf{x}_i,
\end{equation}
whereas $\mathbf{v}_{\text{std}}$ and $\mathbf{x}_{\text{std}}$ are defined as
\begin{equation}
  \mathbf{v}_{\text{std}} = \sqrt{\frac{\sum_{i \in b} w_i \left(\mathbf{v}_i - \mathbf{v}_{\sum}\right)^2}{w_{\sum}}},
  \quad
  \mathbf{x}_{\text{std}} = \sqrt{\frac{\sum_{i \in b} w_i \left(\mathbf{x}_i - \mathbf{x}_{\sum}\right)^2}{w_{\sum}}}.\label{eq:v_x_std}
\end{equation}
In physical terms, $\mathbf{v}_{\sum}$ corresponds to the center-of-mass velocity of the particles in the bin, $\mathbf{x}_{\sum}$ corresponds to the center-of-mass position,
and $\mathbf{v}_{\text{std}}$, $\mathbf{x}_{\text{std}}$ are the per-component variances of the particles' velocities and positions,
respectively. The higher-order moments, however, are not conserved; the reader is referred to~\cite{rjasanow1996stochastic} for a more detailed error analysis of binning-based particle merging schemes.

The necessity of conserving higher-order moments can be motivated by looking at the system of moment equations
obtained from the Boltzmann equation. The Boltzmann equation without external forces reads
\begin{equation}
  \frac{\partial f}{\partial t} + \mathbf{v} \cdot \nabla_{\mathbf{x}} f = Q(f,f),\label{eq:Boltzmann}
\end{equation}
where $Q(f,f)$ is the Boltzmann collision operator.

For simplicity, we write out the system for a one-dimensional distribution $f(x, v)$,
but the same argument holds for the 3-dimensional velocity distributions modelled in the present paper.
By multiplying the Boltzmann equation~(\ref{eq:Boltzmann}) with monomials $1,v,v^2,\ldots$, and integrating over the velocities,
we obtain the following (infinite) system of equations
\begin{align}
  \frac{\partial}{\partial t} M_0 + \frac{\partial}{\partial x} M_1  = S_0,\\
  \frac{\partial}{\partial t} M_1 + \frac{\partial}{\partial x} M_2  = S_1,\nonumber\\
  \frac{\partial}{\partial t} M_2 + \frac{\partial}{\partial x} M_3 = S_2,\nonumber\\
  \ldots\nonumber.
\end{align}
Here $M_n = \int v^n f(v) \mathrm{d} v$, where $f$ is the one-dimensional distribution function, and the collision source term $S_n$ is given by $S_n = \int v^n Q(f,f) \mathrm{d} v$.
Thus, errors induced in the higher-order moments have a cascading effect on the lower-order moments, which are the ones
corresponding to physical quantities of interest such as velocity, temperature, stress tensor, heat flux, etc.

If operator splitting is performed, as is the case in DSMC, the evolution due to advection and due to collisions are treated separately.
Looking at the evolution of a moment at timestep $t$ due to collisions only, we have an initial value problem:
\begin{equation}
  \frac{\partial}{\partial t} M_n = S_n,\:M_n(t_k)=M_n^k.
\end{equation}
However, if merging is performed before the collisions are evaluated, and the merging procedure does not conserve the moment $M_n$, this induces an error both in the evaluation
of the collision operator (as the distribution function is distorted) and in the starting value. This can be formally written as:
\begin{equation}
  \frac{\partial}{\partial t} M_n = S_n + \Delta S_n,\:M_n(t_k)=M_n^k + \Delta M_n^k.
\end{equation}
Here $\Delta S_n$ denotes the change in the evaluation of the collision operator due to the VDF distortion as a result of the merging,
and $\Delta M_n^k$ is the change in the initial value of the moment as a result of the merging. So in this case merging leads to solving a
different evolution equation with a different starting value.
If on the other hand the merging procedure does conserve the value of the moment $M_n$, then the initial value problem takes on the form
\begin{equation}
  \frac{\partial}{\partial t} M_n = S_n + \Delta S_n,\:M_n(t_k)=M_n^k.
\end{equation}
Here $\Delta' S_n$ also denotes the change in the evaluation of the collision operator due to the VDF distortion as a result of the merging.
However, it can be reasonably assumed that if merging conserves higher-order moments of the distribution function, the induced error
in the evaluation of collisions is smaller compared to an algorithm that does not conserve this information.

The conservation of spatial moments has a less direct effect on the evolution of the system,
but affects the fluxes of the particles through the cell boundaries,
as well as charge conservation in particle-in-cell simulations~\cite{martin2016octree}.
Thus, preservation of moments during merging is highly important for reducing the merging-induced error in particle simulations.

\section{Moment-preserving merging}
\label{subsec:moment}

We are therefore interested in the conservation of $N_v$ velocity and $N_x$ spatial moments, which we
denote as $M_{v,j}$ and $M_{x,j}$ correspondingly. Each is defined by a
multi-index set $\mathcal{I}_v = \{(m_{v_{x},j},m_{v_{y},j},m_{v_{z},j}),\: j=1,\ldots,N_v\}$,
$\mathcal{I}_x = \{(m_{x_{x},j},m_{x_{y},j},m_{x_{z},j}),\: j=1,\ldots,N_x\}$:
\begin{equation}
  M_{v,j} = \sum_{i \in \mathcal{P}} w_i v_{i,x}^{m_{v_{x},j}}v_{i,y}^{m_{v_{y},j}}v_{i,z}^{m_{v_{z},j}},\:j=1,\ldots,N_v \label{eq:def-vel-moment}
\end{equation}
\begin{equation}
  M_{x,j} = \sum_{i \in \mathcal{P}} w_i v_{i,x}^{m_{x_{x},j}}v_{i,y}^{m_{x_{y},j}}v_{i,z}^{m_{x_{z},j}},\:j=1,\ldots,N_x.
\end{equation}
Here $\sum_{i \in \mathcal{P}}$ denotes the summation over the original set of pre-merge particles $\mathcal{P}$.
We can alternately write the definitions of the moments in a matrix-vector product form:
\begin{equation}
  \mathbf{V} \mathbf{w} = \mathbf{m},\label{eq:linear-eq-matrix}
\end{equation}
where $\mathbf{V} \in \mathbb{R}^{(N_v+N_x)\times N}$ is given by
\begin{equation}
 \mathbf{V} = \begin{pmatrix}
v_{1,x}^{m_{v_{x},1}}v_{1,y}^{m_{v_{y},1}}v_{1,z}^{m_{v_{z},1}}  & \ldots & v_{N,x}^{m_{v_{x},1}}v_{N,y}^{m_{v_{y},1}}v_{N,z}^{m_{v_{z},1}}\\
& \ldots & \\
v_{1,x}^{m_{v_{x},N_v}}v_{1,y}^{m_{v_{y},N_v}}v_{1,z}^{m_{v_{z},N_v}}  & \ldots & v_{N,x}^{m_{v_{x},N_v}}v_{N,y}^{m_{v_{y},N_v}}v_{N,z}^{m_{v_{z},N_v}}\\
\\
x_{1,x}^{m_{x_{x},1}}x_{1,y}^{m_{x_{y},1}}x_{1,z}^{m_{x_{z},1}}  & \ldots & x_{N,x}^{m_{x_{x},1}}x_{N,y}^{m_{x_{y},1}}x_{N,z}^{m_{x_{z},1}}\\
& \ldots & \\
x_{1,x}^{m_{x_{x},N_x}}x_{1,y}^{m_{x_{y},N_x}}x_{1,z}^{m_{x_{z},N_x}}  & \ldots & x_{N,x}^{m_{x_{x},N_x}}x_{N,y}^{m_{x_{y},N_x}}x_{N,z}^{m_{x_{z},N_x}}
\end{pmatrix},\label{eq:moment-matrix-def}
\end{equation}
and the vectors $\mathbf{w} \in \mathbb{R}^N$, $\mathbf{m} \in \mathbb{R}^{N_v+N_x}$ are given by
\begin{equation}
  \mathbf{w} = \begin{pmatrix}
w_1 \\
\vdots \\
w_N
\end{pmatrix},
\:
\mathbf{m} = \begin{pmatrix}
m_{v,1} \\
\vdots \\
m_{v,N_v}\\
m_{x,1}\\
\vdots \\
m_{x,N_x}
\end{pmatrix}.\label{eq:moment-rhs-def}
\end{equation}

A moment-conserving $N:M$ merging procedure can then be formulated as follows: find $\mathbf{V}$, $\mathbf{w}$ such that~(\ref{eq:moment-matrix-def}) holds, $\mathbf{w} \geq 0$, and $\mathbf{w}$ has only $M<N$ non-zero entries, that is, $\mathbf{w}$ is sparse. A sparse
weight vector directly corresponds to merging, as the particles with post-merge weights of 0 can simply be discarded from the simulation.
Although at first glance such posing of the problem is similar to that arising in \textit{sparse dictionary learning}~\cite{kreutz2003dictionary},
in the case of particle merging the rows of matrix $\mathbf{V}$ are not independent, as $\mathbf{V}$ is a concatenation of multi-variate
Vandermonde matrices of the velocities and positions of the pre-merge particles.
An additional challenges is posed by the fact that optimizing for the entries of $\mathbf{V}$,
i.e. moments of the post-merge particle positions and velocities, may lead to out-of-domain particles, as discussed above.
This general setting is also highly non-linear and non-convex~\cite{bohmer2020entropic}, and is exacerbated by the fact that Vandermonde matrices tend to be badly conditioned~\cite{pan2016bad}.

To avoid the issues of out-of-domain particles and simplify the problem setting,
we can instead consider the sparsity-inducing optimization problem
\begin{align}
  \min_{w_i \geq 0} \quad &   \lambda ||\mathbf{w}||_0 \\
  \textrm{s.t.}  \quad & \mathbf{V} \mathbf{w} = \mathbf{m},\label{eq:}
\end{align}
where the $l_0$ pseudo-norm $||\mathbf{w}||_0$ is the number of non-zero entries of $\mathbf{w}$. This approach corresponds to choosing new weights but re-using the velocities and positions of the pre-merge particles (since $\mathbf{V}$ remains fixed),
this guaranteeing the preservation of any spatial and phase space bounds on their positions and velocities.

However, as the $l_0$ optimization problem is NP-hard, the $l_1$ norm is often optimized for instead, as it also leads to sparse solutions.
In the case of $w_i \geq 0$ the $l_1$ norm however cannot be optimized, as $||\mathbf{w}||_1 = \sum_{i \in \mathcal{P}} |w_i| = \sum_{i \in \mathcal{P}} w_i = w_{\sum}$, that is, the norm is simply the number density of the pre-merge particles, which remains constant throughout the merging procedure. Our recent work on multi-dimensional moment closures~\cite{oblapenko2025sparse} can be viewed as a particle merging algorithm, where
sparsity induced by optimizing for the sum of inter-particle distances in velocity space and subsequently
replacing particles that are close together with a single particle.

Nevertheless, the general idea of re-using a subset of the pre-merge particles' positions and velocities and finding
new weights for these particles has found traction in recent years.
Faghihi et al.~\cite{faghihi2020moment} propose a re-sampling based approach procedure for moment-preserving particle reduction.
The phase and physical space are discretized into $d_v + d_x$-dimensional bins, and in each bin, a new set of particle positions and
velocities is sampled, either
as a sample from the existing pre-merge particles, or from a uniform distribution on the phase space bin.
The weights of these particles are then found as a constrained minimization problem, where the target function is the variance of the weights.
Since the moments depend linearly the weights, this leads to a quadratic optimization problem with linear constraints, which can be efficiently
solved. Gonoskov proposed an iterative procedure for moment-preserving merging that deletes one particle per step~\cite{gonoskov2022agnostic}, this
has been later coupled with the octree velocity space-binning approach to reduce the error induced by the merging~\cite{huerta2024situ}. An alternative iterative approach to 
moment-preserving system reduction with Vandermonde matrices has recently been developed in~\cite{bvelik2025efficient}.

The recent work of Goeckner et al.~\cite{goeckner2025generalized} proposes constructing an invertible square matrix $\mathbf{V}$ to use to
directly solve~(\ref{eq:linear-eq-matrix}) such that the solution is non-negative. This is carried out analytically for velocity moments up to total
order of 3, i.e. $\{(m_{v,x},m_{v,y},m_{v,z}) | m_{v,x} + m_{v,y} + m_{v,z} \leq 3 \}$. 
Finally, the roulette merge~\cite{watrous2023improvements}
uses a random deletion of $N-M$ particles and re-weighting of the remaining ones so as to conserve number density;
it ensures the conservation of any moment of the distribution on-average, that is, in the limit of infinitely many merges. However,
the stochastic nature of the approach is an additional source of noise in the simulation. Moreover, it has been observed in collisional simulations that a stochastic deletion of particles may lead to such a re-weighting that results in certain particles having very large computational weights, leading to inefficient evaluation of collisions~\cite{schmidt2000new}.

\subsection{Non-negative least squares-based merging}
\label{subsec:NNLS}
Recently it has been proposed~\cite{oblapenko2024non}  to treat the underdetermined set of equations (\ref{eq:linear-eq-matrix})
with the constraint $w_i \geq 0$ directly as a non-negative least-squares (NNLS) system~\cite{lawson1995solving}.
Underdetermined NNLS systems are known to result in sparse solutions~\cite{slawski2013non,dessole2023lawson}, so if the total number of velocity and spatial moment constraints
$M_v + M_x$ is less than the pre-merge number of particles $N$, one expects to obtain a sparse solution vector $\mathbf{w}$ with $M_v + M_x$ non-zero entries. Discarding the zero weights, one can thus obtain $M_v + M_x$ post-merge particles, with their velocities and positions taken from the
original set of velocities and positions of the pre-merge particles, by using the indices of the non-zero weights to look up the corresponding
properties of the pre-merge particles.

In this section, we discuss the adjustments to the system (\ref{eq:linear-eq-matrix}) that improve the numerical stability of the algorithm,
and extensions of the NNLS approach to rate-conserving merging in plasmas.
The NNLS algorithm used in the present work is an implementation of the algorithm of Lawson and Hanson~\cite{lawson1995solving} , which for a system $\mathbf{V}\mathbf{w}=\mathbf{m}$, $\mathbf{w} \geq 0$ is given by Alg.~\ref{alg:nnls}.

\begin{algorithm}[h]\label{alg:nnls}
\caption{Lawson--Hanson NNLS algorithm for solving a linear system $\mathbf{V}\mathbf{w}=\mathbf{m}$ with constraint $\mathbf{w} \geq 0$}
\DontPrintSemicolon
$P \gets \emptyset$, $Z \gets \{1,\ldots,N\}$, $\mathbf{w} \in \mathbb{R}^{N} \gets 0$, $\mathbf{x} \in \mathbb{R}^{N} \gets -\mathbf{V}^T \mathbf{m}$\;
\While{$Z \neq \emptyset$ \textbf{and} $\max(\mathbf{x}) > 0$}{
    $\tau \gets \arg\max_i \mathbf{x}_i$\;
    move $\tau$ from $Z$ to $P$\;
    $\mathbf{z}_P \gets $ LS solution of $\mathbf{V}_P \mathbf{w} = \mathbf{m}$ \; 
    $\mathbf{z}_Z \gets 0$\;
    \While{$\min(z_P) \le 0$}{
        $Q \gets P \cap \{\, i : \mathbf{z}_i \le 0\,\}$\;
        $\alpha \gets \displaystyle\min_{i \in Q} \frac{\mathbf{w}_i}{\mathbf{w}_i - \mathbf{z}_i}$\;
        $\mathbf{w} \gets \mathbf{w} + \alpha (\mathbf{z} - \mathbf{w})$\;
        move $\{\, i : i \in P,\; \mathbf{w}_i \le 0 \,\}$ from $P$ to $Z$\;
        $\mathbf{z}_P \gets $ LS solution of $\mathbf{V}_P \mathbf{w} = \mathbf{m}$, $\mathbf{z}_Z \gets 0$\ \;
    }
    $\mathbf{w} \gets \mathbf{z}$, 
    $\mathbf{x} \gets \mathbf{V}^T (\mathbf{V} \mathbf{w}  - \mathbf{m})$\;
}
$\mathbf{w}^\star \gets \mathbf{w}$.\;
\end{algorithm}
The vector $\mathbf{w}^\star$ obtained at the termination of the outer loop is the solution of the NNLS system.

Here the vector $\mathbf{z}$ is a $N$-dimensional vector, whose components
are defined through the sub-vectors $\mathbf{z}_P$, $\mathbf{z}_Z$, i.e. through the sets of indices $P$, $Z$.
$P$ holds the indices of non-zero entries of $\mathbf{z}$, whereas $Z$ holds the indices of entries of $\mathbf{z}$ equal to zero.
$\mathbf{V}_P$ is a matrix of the same size as $\mathbf{V}$, with columns whose indices are in $P$ taken equal to the columns of $\mathbf{V}$, and as columns filled with
zeros otherwise. ``LS solution of $\mathbf{V}_P \mathbf{w} = \mathbf{m}$'' denotes the least-squares solution of the underdetermined system computed
using QR factorization. This solution determines only the components of $\mathbf{z}_P$, and the components of $\mathbf{z}_Z$ are set to 0.

As mentioned previously, the conditioning of the matrix $\mathbf{V}$ appearing in the NNLS system is expected to be poor.
To improve the convergence rate, we solve the following scaled system:
\begin{align}
  \mathbf{R}\hat{\mathbf{V}}\mathbf{S}\mathbf{x} & = \hat{\mathbf{m}},\nonumber\\
  \mathbf{x} & \geq 0, \label{eq:nnls-scaled}
\end{align}
where the elements of $\hat{\mathbf{M}}$ are given by
\begin{equation}
  \hat{m}_{v,j} = \frac{1}{w_{\sum} v_{\mathrm{ref},x}^{m_{v_{x},j}} v_{\mathrm{ref},y}^{m_{v_{y},j}} v_{\mathrm{ref},z}^{m_{v_{z},j}} } M_{v,j},\:j=1,\ldots,N_v \label{eq:scaling-m-v}
\end{equation}
\begin{equation}
  \hat{m}_{x,j} = \frac{1}{w_{\sum} x_{\mathrm{ref},x}^{m_{x_{x},j}} x_{\mathrm{ref},y}^{m_{x_{y},j}} x_{\mathrm{ref},z}^{m_{x_{z},j}} } M_{x,j},\:j=1,\ldots,N_x,  \label{eq:scaling-m-x}
\end{equation}
$\mathbf{v}_{\mathrm{ref}}$ and $\mathbf{x}_{\mathrm{ref}}$ are reference velocity and position vectors, and the diagonal matrix $\mathbf{S} \in \mathbb{R}^{N\times N}$ is defined as
\begin{equation}
  \mathbf{S} = \mathrm{diag}\left(\frac{1}{||(\mathbf{R}\hat{\mathbf{V}})_1||_2}, \ldots, \frac{1}{||(\mathbf{R}\hat{\mathbf{V}})_N||_2} \right).
\label{eq:moment-matrix-scaling-S-def}
\end{equation}
Here $||(\mathbf{R}\hat{\mathbf{V}})_i||_2$ is the $L_2$ norm of column of $i$ of the matrix $\mathbf{R}\hat{\mathbf{V}}$, where $\hat{\mathbf{V}}$ is computed similarly to
(\ref{eq:moment-matrix-def}), but using offset velocities $\hat{\mathbf{v}}_i$ and positions $\hat{\mathbf{x}}_i$ instead of $\mathbf{v}_i$, $\mathbf{x}_i$:
\begin{equation}
  \hat{\mathbf{v}}_i = \mathbf{v}_i - \mathbf{v}_{\sum},\quad
  \hat{\mathbf{x}}_i = \mathbf{x}_i - \mathbf{x}_{\sum}.\label{eq:scaling-R-def}
\end{equation}

Finally, the diagonal matrix $\mathbf{R} \in \mathbb{R}^{(N_v+N_x) \times (N_v+N_x)}$ is a scaling of its rows, defined as
\begin{equation}
 \mathbf{R} = 
 \begin{pmatrix}
   \mathbf{R}_v & \mathbf{0}_{N_v \times N_x} \\
   \mathbf{0}_{N_x \times N_v} & \mathbf{R}_x 
 \end{pmatrix},
\label{eq:moment-matrix-scaling-R-def}
\end{equation}
\begin{equation}
  \mathbf{R}_v = 
  \mathrm{diag}\left(\frac{1}{v_{\mathrm{ref},x}^{m_{v_{x},1}} v_{\mathrm{ref},y}^{m_{v_{y},1}} v_{\mathrm{ref},z}^{m_{v_{z},1}}}, \ldots,
  \frac{1}{v_{\mathrm{ref},x}^{m_{v_{x},N_v}} v_{\mathrm{ref},y}^{m_{v_{y},N_v}} v_{\mathrm{ref},z}^{m_{v_{z},N_v}}} \right),
\end{equation}
\begin{equation}
  \mathbf{R}_x = 
  \mathrm{diag}\left(\frac{1}{x_{\mathrm{ref},x}^{m_{x_{x},1}}x_{\mathrm{ref},y}^{m_{x_{y},1}}x_{\mathrm{ref},z}^{m_{x_{z},1}}}, \ldots,
  \frac{1}{x_{\mathrm{ref},x}^{m_{x_{x},N_x}}x_{\mathrm{ref},y}^{m_{x_{y},N_x}}x_{\mathrm{ref},z}^{m_{x_{z},N_x}}} \right).
\end{equation}
Once a solution $\mathbf{x}$ of (\ref{eq:nnls-scaled}) is found, the vector of post-merge weights of the particles $\mathbf{w}$ can be
obtained as $w_{\sum} \mathbf{S} \mathbf{x}$.
For example, the reference velocity and position vectors used for the scaling can be computed as variances of the pre-merge
particles' positions and velocities, correspondingly (see Eqn.~(\ref{eq:v_x_std})):
\begin{equation}
  \mathbf{x}_\mathrm{ref} = \mathbf{x}_\mathrm{std},\quad \mathbf{v}_\mathrm{ref} = \mathbf{v}_\mathrm{std}.
\end{equation}

\begin{algorithm}[H]\label{alg:nnls-merge}
\caption{NNLS particle merging algorithm}
\DontPrintSemicolon

$\hat{\mathbf{V}} \in \mathbb{R}^{(N_v + N_x)\times N} \gets 0$, $\hat{\mathbf{w}} \in \mathbb{R}^{N} \gets 0$, $\hat{\mathbf{m}} \in \mathbb{R}^{N_v + N_x} \gets 0$,  ${\mathbf{s}} \in \mathbb{R}^{N} \gets 0$, $P \gets \emptyset$\;

$w_{\sum} \gets \sum_i w_i$, ${\mathbf{v}}_{\sum} \gets w_{\sum}^{-1} \sum_i w_i {\mathbf{v}}_i$,
${\mathbf{x}}_{\sum} \gets w_{\sum}^{-1} \sum_i w_i {\mathbf{x}}_i$\;

$\hat{w}_i \gets w_{\sum}^{-1}w_i$, $\hat{\mathbf{v}}_i \gets {\mathbf{v}}_i - {\mathbf{v}}_{\sum}$, $\hat{\mathbf{x}}_i \gets {\mathbf{x}}_i - {\mathbf{x}}_{\sum}$,\: $i=1,\ldots,N$\;

$\mathbf{v}_{\mathrm{ref}} \gets \sqrt{w_{\sum}^{-1}\sum_i w_i \hat{\mathbf{v}}_i^2}$, $\mathbf{x}_{\mathrm{ref}} \gets \sqrt{w_{\sum}^{-1}\sum_i w_i \hat{\mathbf{x}}_i^2}$ // compute values required for scaling \;

$\hat{\mathbf{v}}_i \gets \mathbf{v}_{\mathrm{ref}}^{-1} \hat{\mathbf{v}}_i$, $\hat{\mathbf{x}}_i \gets \mathbf{x}_{\mathrm{ref}}^{-1} \hat{\mathbf{x}}_i$,\: $i=1,\ldots,N$  // equivalent to left-scaling with $\mathbf{R}$ defined by (\ref{eq:moment-matrix-scaling-R-def}) \;

$\hat{\mathbf{V}}_{j,i} \gets \hat{\mathbf{v}}_{i,x}^{m_{v_x,j}} \hat{\mathbf{v}}_{i,y}^{m_{v_y,j}} \hat{\mathbf{v}}_{i,z}^{m_{v_z,j}}$,\: $i=1,\ldots,N$,  \: $j=1,\ldots,N_v$\;

$\hat{\mathbf{V}}_{N_v+j,i} \gets \hat{\mathbf{x}}_{i,x}^{m_{x_x,j}} \hat{\mathbf{x}}_{i,y}^{m_{x_y,j}} \hat{\mathbf{x}}_{i,z}^{m_{x_z,j}}$,\: $i=1,\ldots,N$,  \: $j=1,\ldots,N_x$\;

$\hat{\mathbf{m}}_j \gets \sum_i \hat{w}_i \hat{\mathbf{V}}_{j,i}$,  \: $j=1,\ldots,N_x$\;

$s_i \gets \left(\sum_j \hat{\mathbf{V}}_{j,i}^2  \right)^{-1/2}$,\: $i=1,\ldots,N$  // compute $\mathbf{S}$ defined by (\ref{eq:moment-matrix-scaling-S-def}) \;

$\hat{\mathbf{V}}_{j,i} \gets s_i \hat{\mathbf{V}}_{j,i}$,\: $i=1,\ldots,N$, $j=1,\ldots,N_v + N_x$ // right-scaling with (\ref{eq:moment-matrix-scaling-S-def}) \; 

Solve NNLS system $\hat{\mathbf{V}}_{j,i} \mathbf{x} = \hat{\mathbf{m}}_j$, $\mathbf{x} \geq 0$ using Algorithm~\ref{alg:nnls}\;

$\hat{\mathbf{w}} \gets \mathbf{s} \odot \mathbf{x}$ // obtain weights from NNLS solution \;

$P \gets \{\, i : \hat{w}_i \geq \varepsilon\,\}$

\For{$k=1,\ldots,\mathrm{card}(P)$}{
 particle $k$ $\gets$ $\left(w_{\sum} \hat{w}_{P_k}, \: \mathbf{v}_{P_k}, \: \mathbf{x}_{P_k} \right)$ // post-merge particles
}
\For{$k=\mathrm{card}(P)+1,\ldots,N$}{
 particle $k$ is deleted
}
\end{algorithm}
Algorithm \ref{alg:nnls-merge} outlines the steps of the NNLS-based merging algorithm.
Here $\mathrm{card}(P)$ denotes the size of set $P$, $\mathbf{a}^{-1}$ denotes a component-wise inverse of a vector $\mathbf{a}$, and $\odot$ denotes the component-wise product of two vectors. As solutions of the NNLS problem may be not exactly zero due to round-off errors, a tolerance $\varepsilon > 0$ is specified, with relative weights $\hat{w}_i < \varepsilon$ being deleted.
This concludes the description of the proposed moment-preserving NNLS-based merging algorithm.

\subsection{Rate-preserving merging}
\label{subsec:NNLSrp}
The distortion introduced into the velocity distribution function during merging can have a strong impact on the reaction rate coefficients~\cite{oblapenko2022improving}.
As such, a merging procedure that conserves reaction rate coefficients would improve simulation accuracy. In a particle-based discretization,
the expression for a reaction rate coefficients for process $r$ for particles of species $s_1$ and $s_2$ reads
\begin{equation}
  k_r = \frac{1}{\sum_i w^{(s_1)}_i \sum_k w^{(s_2)}_k}\sum_{i=1}^{N^{(s_1)}} \sum_{k=1}^{N^{(s_2)}} w^{(s_1)}_i w^{(s_2)}_k |\mathbf{v}^{(s_1)}_i - \mathbf{v}^{(s_2)}_k| \sigma_r \left(|\mathbf{v}^{(s_1)}_i - \mathbf{v}^{(s_2)}_k|\right).\label{eq:kr}
\end{equation}
Here superscripts $s_1$, $s_2$ denote the chemical species of the particles, and $\sigma_r$ is the total cross-section of the process, dependent on the magnitude of the relative velocity of
the colliding particles.
It can be seen that conserving the rate when merging particles requires consideration of all possible collision pairs, a procedure with a cost of $\mathcal{O}\left(N^{(s_1)} N^{(s_2)}\right)$.

For the case of NNLS merging, the rate conservation for the case of distinct species $s_1 \neq s_2$ can be added via adding rows to the matrix $\mathbf{V}$ and entries to the right-hand side vector $\mathbf{m}$.
Namely, assuming we are merging particles of species $s_1$ and particles of species $s_2$ remain unaffected, we introduce the matrix $\mathbf{V}_{RP} \in \mathbb{R}^{(N_v + N_x + N_r)\times N^{(s_1)}}$:
\begin{equation}
  \mathbf{V}_{RP} = \begin{pmatrix}
    \mathbf{V} \\
    \mathbf{V}_{R}
  \end{pmatrix},
\end{equation}
where $\mathbf{V}$ is given by~(\ref{eq:moment-matrix-def}), $N_r$ is the number of rates being conserved, and the matrix $\mathbf{V}_{R} \in \mathbb{R}^{N_r \times N^{(s_1)}}$ is defined as
\begin{equation}
  \mathbf{V}_{R} = \begin{pmatrix}
    \sum_{k=1}^{N^{(s_2)}} w^{(s_2)}_k g_{1,k}^{(s_1,s_2)} \sigma_{1} \left(g_{1,k}^{(s_1,s_2)}\right) & \ldots & \sum_{k=1}^{N^{(s_2)}} w^{(s_2)}_k g_{N^{(s_1)},k}^{(s_1,s_2)} \sigma_{1} \left(g_{N^{(s_1)},k}^{(s_1,s_2)}\right)\\
& \ldots & \\
\sum_{k=1}^{N^{(s_2)}} w^{(s_2)}_k g_{1,k}^{(s_1,s_2)} \sigma_{N_r} \left(g_{1,k}^{(s_1,s_2)}\right) & \ldots & \sum_{k=1}^{N^{(s_2)}} w^{(s_2)}_k g_{N^{(s_1)},k}^{(s_1,s_2)} \sigma_{N_r} \left(g_{N^{(s_1)},k}^{(s_1,s_2)}\right)
  \end{pmatrix}.
\end{equation}
Here we use notation $g_{i,k}^{(s_1,s_2)} = |\mathbf{v}^{(s_1)}_i - \mathbf{v}^{(s_2)}_k|$, and $\sigma_1,\ldots, \sigma_{N_r}$ are the total cross-sections of the different rates.
For the right-hand side, we introduce the vector $\mathbf{m}_{RP} \in \mathbb{R}^{N_v + N_x + N_r}$:
\begin{equation}
  \mathbf{m}_{RP} = \begin{pmatrix}
    \mathbf{m} \\
    \mathbf{m}_{R}
  \end{pmatrix},
\end{equation}
where $\mathbf{m}$ is given by~(\ref{eq:moment-rhs-def}), and the vector $\mathbf{m}_{R} \in \mathbb{R}^{N_r \times N^{(s_1)}}$ is defined as
\begin{equation}
  \mathbf{m}_{R} = \begin{pmatrix}
    \sum_{i=1}^{N^{(s_1)}} \sum_{k=1}^{N^{(s_2)}} w^{(s_1)}_i w^{(s_2)}_k g_{i,k}^{(s_1,s_2)} \sigma_1 \left(g_{i,k}^{(s_1,s_2)}\right)\\
    \vdots\\
    \sum_{i=1}^{N^{(s_1)}} \sum_{k=1}^{N^{(s_2)}} w^{(s_1)}_i w^{(s_2)}_k g_{i,k}^{(s_1,s_2)} \sigma_{N_r} \left(g_{i,k}^{(s_1,s_2)}\right)
  \end{pmatrix}.
\end{equation}
Then the NNLS system $\mathbf{V}_{RP} \mathbf{w} = \mathbf{m}_{RP},\:\mathbf{w} \geq 0$ corresponds to conservation of the rates of process $r=1,\ldots,N_r$. As the rates appearing $\mathbf{m}_{R}$ are not
scaled by the number densities of species $s_1$, $s_2$, this assumes that number density is conserved in the merging. It should be noted that while the species $s_1$ is fixed, as it is the species of the particles
being merged, species $s_2$ can vary across the rows, corresponding to conservation of rates of collisional processes for interactions of particles of species $s_1$ with different chemical species.

The solution algorithm is almost identical to Algorithm~\ref{alg:nnls-merge}, except that
\begin{enumerate}
  \item The entries of $\mathbf{V}_{R}$ and $\mathbf{m}_{R}$ are computed using the non-shifted non-scaled particles' velocities,
as the computation of rates and scaling do not commute.
  \item Each row $r$ of $\mathbf{V}_{R}$ and corresponding entry of $\mathbf{m}_{R}$ is scaled by \newline $\sum_i w^{(s_1)}_i \sum_k w^{(s_2)} k_{\mathrm{ref},r}$, where $k_{\mathrm{ref},r}$ is a reference reaction rate coefficient for process $r$.
\end{enumerate}

It should be emphasized that if the goal is the preservation of a certain process rate, the corresponding preservation constraints should be applied both when merging particles of species $s_1$
as well as of species $s_2$. For example, if particles of species $s_1$ are merged first, then the rate preservation constraints for the merging of particles of species $s_2$ should be based on the new post-merge particles of species $s_1$. This hinders species-wise parallelization of the merging process, adds additional computational complexity (as the pair-wise contributions to the rates need to be computed twice),
and precludes the use of non-rate-preserving merging algorithms for either of the species.

\subsection{Approximate rate-preserving merging}
However, a simplified approximate rate preservation condition can be derived for electron-neutral collision rates for plasma simulations. Assuming without loss of generality that species $s_1$ corresponds to the
electrons and species $s_2$ to the neutrals, it is often the case that the velocities of the electrons are several orders of magnitude larger than those of the neutrals.
It can then be written that
\begin{equation}
  g_{i,k}^{(s_1,s_2)} = |\mathbf{v}^{(s_1)}_i - \mathbf{v}^{(s_2)}_k|\approx |\mathbf{v}^{(s_1)}_i|.
\end{equation}
This allows to derive an expression for an approximate process rate $k_{r,a}$ that is independent of the properties of the particles of species $s_2$:
\begin{equation}
  k_{r,a} = \frac{1}{\sum_i w^{(s_1)}_i} \sum_{i=1}^{N^{(s_1)}} w^{(s_1)}_i |\mathbf{v}^{(s_1)}_i| \sigma_r \left(|\mathbf{v}^{(s_1)}_i|\right).\label{eq:kr_approx}
\end{equation}
The conservation of these approximate rates in the NNLS merging framework is then ensured by solving the system $\mathbf{V}_{ARP} \mathbf{w} = \mathbf{m}_{ARP},\:\mathbf{w} \geq 0$,
where 
\begin{equation}
  \mathbf{V}_{ARP} = \begin{pmatrix}
    \mathbf{V} \\
    \mathbf{V}_{AR}
  \end{pmatrix},
\end{equation}
\begin{equation}
  \mathbf{V}_{AR} = \begin{pmatrix}
    |\mathbf{v}^{(s_1)}_1| \sigma_{1} \left(|\mathbf{v}^{(s_1)}_1|\right) & \ldots & |\mathbf{v}^{(s_1)}_{N^{(s_1)}}| \sigma_{1} \left(|\mathbf{v}^{(s_1)}_{N^{(s_1)}}| \right)\\
& \ldots & \\
|\mathbf{v}^{(s_1)}_1| \sigma_{N_r} \left(|\mathbf{v}^{(s_1)}_1|\right) & \ldots & |\mathbf{v}^{(s_1)}_{N^{(s_1)}}|  \sigma_{N_r} \left(|\mathbf{v}^{(s_1)}_{N^{(s_1)}}| \right)
  \end{pmatrix},
\end{equation}
\begin{equation}
  \mathbf{m}_{ARP} = \begin{pmatrix}
    \mathbf{m} \\
    \mathbf{m}_{AR}
  \end{pmatrix},
\end{equation}
\begin{equation}
  \mathbf{m}_{AR} = \begin{pmatrix}
    \sum_{i=1}^{N^{(s_1)}} w^{(s_1)}_i  |\mathbf{v}^{(s_1)}_i| \sigma_1 \left(|\mathbf{v}^{(s_1)}_i|\right)\\
    \vdots\\
    \sum_{i=1}^{N^{(s_1)}} w^{(s_1)}_i  |\mathbf{v}^{(s_1)}_i| \sigma_{N_r} \left(|\mathbf{v}^{(s_1)}_i|\right)
  \end{pmatrix}.
\end{equation}
The scaling for each row $r$ of $\mathbf{V}_{AR}$ and corresponding entry of $\mathbf{m}_{AR}$ in this case is given by $\sum_i w^{(s_1)}_i k_{\mathrm{ref},r}$.
This concludes the discussion of the rate-preserving and approximate rate-preserving versions of the NNLS-based merging algorithm.

\section{Numerical results}
\label{sec:numerics}
The merging approaches were implemented in the open-source variable-weight DSMC code Merzbild.jl~\cite{oblapenko2024merzbild} developed by the first author. The results were obtained using versions 0.7.8 and 0.8.2 of the code. The workflow required to reproduce the results presented below is described in \texttt{PAPER\_REPRODUCIBILITY.md} in the root of the repository of Merzbild.jl.
The simulation data produced by the solver is also available 
via Zenodo~\cite{oblapenko2026researchdata}.

We compare results of the NNLS-based
merging with those obtained with the octree binning approach proposed in~\cite{martin2016octree}, as the latter is a well-established, computationally efficient, and adaptive algorithm.
\blue{We use the $N:2$ merging strategy in each octree bin as described in~\cite{martin2016octree}, i.e. only density, momentum, and energy (in each direction) are conserved; higher-order merging strategies such as that described in~\cite{lama2020higher}
are not considered in the present work.}
\blue{For the NNLS merging, we always consider conservation of all mixed velocity moments up to order $L$, i.e. the powers $m_{r,j},\:r=x,y,z$ in~(\ref{eq:def-vel-moment}) are defined
by the relation $0 \leq \sum_{r=x,y,z} m_{r,j} \leq L$, any other additional quantities conserved (i.e. rates, spatial moments) are mentioned separately in the relevant sections.}

\subsection{Merging of sampled equilibrium distribution}
We first consider the case of merging of particles sampled from a Maxwell--Boltzmann distribution, similar to the case considered in~\cite{hong2024improved}.
We sample 500 particles from the distribution, merge them, and evaluate the tail functions $F(v)$, defined as
\begin{equation}
  F(v) = \frac{1}{w_{\Sigma}}\sum_{i: ||\mathbf{v}_i||_2 \geq v}  w_i. 
\end{equation}
That is, the tail function is the fraction of the distribution with speeds larger than $v$. We evaluate $F(v)$ at values of $v=250$~m/s and $v=750$~m/s.
We also consider the standard deviation of the post-merge particle weights $\sigma_w$; as these can potentially vary over several orders of magnitude, the standard deviation of the logarithm of the post-merge particle weights $\sigma_{\ln w}$ is also computed. Finally, the ratio of the largest post-merge particle weight $w_{\mathrm{max}}$ to the smallest non-zero post-merge particle weight $w_{\mathrm{min}}$ is also evaluated.

These quantities are computed by sampling 500 particles of an argon gas in a 1~m$^3$ box with a total number density of 1~m$^{-3}$
from a stationary Maxwell--Boltzmann distribution at a temperature of 300~K. These are then immediately merged and the quantities
described above evaluated. These procedure is repeated 100000 times and the mean values of the aforementioned quantities of interest computed. We investigate two different sampling strategies, described in detail below.
\blue{For the octree merging, we use target numbers of particles ranging from $96$ to $220$, and for the NNLS merging, we vary the highest total order $L$ of the conserved velocity moments from 4 to $9$.}

\subsubsection{Equal-weight samples}\label{sec:equal-weight}
First, we consider sampling of equal-weight particles. This is performed in the standard fashion deployed in DSMC codes~\cite{DSMC_Bird}. Even if the simulation starts out with equal-weight particles, the merging
procedure immediately leads to a variation in the weights.

\begin{figure}[t]
  \centering
  \includegraphics[width=0.98\textwidth]{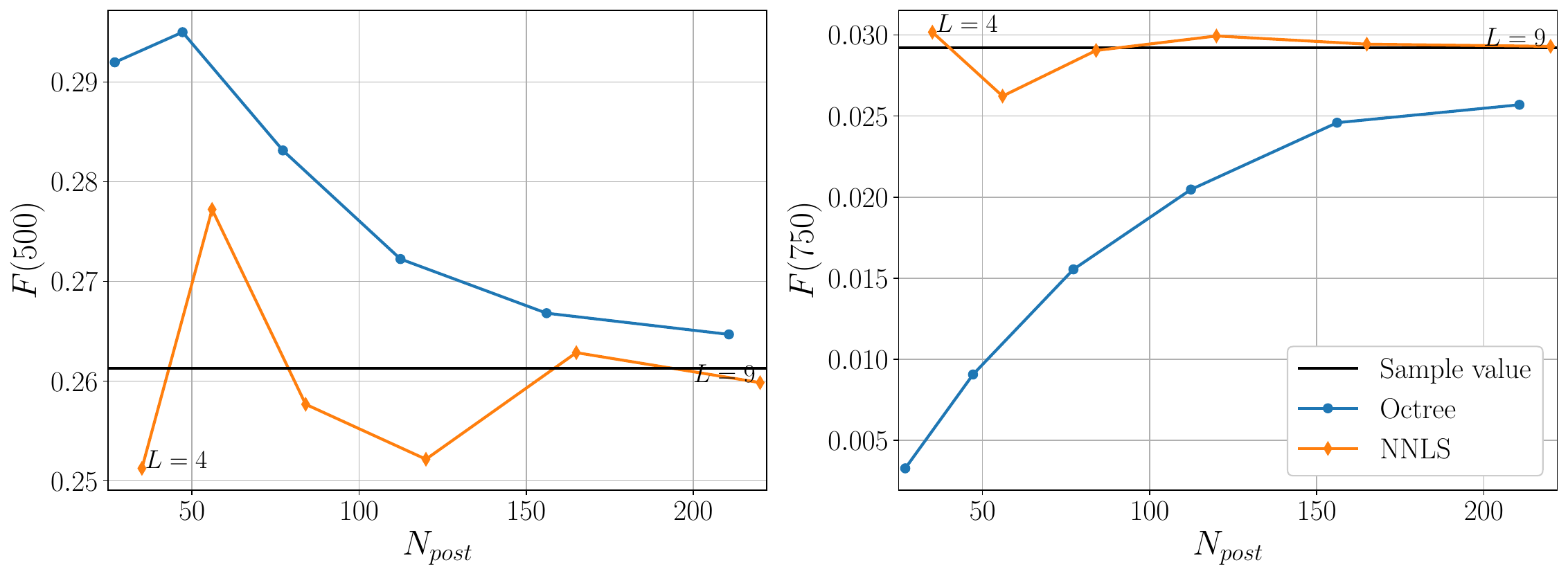}
  \caption{Values of the tail functions $F(500)$ (left) and $F(750)$ (right) as a number of post-merge particles, pre-merge particles with equal weights.}\label{fig:merging-tf-equalweight}
  \end{figure}
Figure~\ref{fig:merging-tf-equalweight} shows the values of the tail functions $F(500)$ and $F(750)$ as a number of post-merge particles $N_{\mathrm{post}}$, with the initial value of the ensemble shown in black.
We see that for the high-speed tail $F(750)$, the NNLS merging preserves the number density in the tail much more accurately than the octree-based merging algorithm. For the tail function $F(500)$, the behaviour
of the NNLS merging algorithm is less clear. Since the merging is based on the moments, the speed distribution at lower speeds might be less well-preserved, as the low-speed particles contribute less
to the moments. For the octree merging, due to its iterative sub-division of the velocity space, the accuracy of the representation of the distribution function increases with an increasing number of
target post-merge particles, as the velocity bins are refined and become smaller, thus avoiding merging of particles that lie far from each in phase space.

  \begin{figure}[t]
    \centering
    \includegraphics[width=0.98\textwidth]{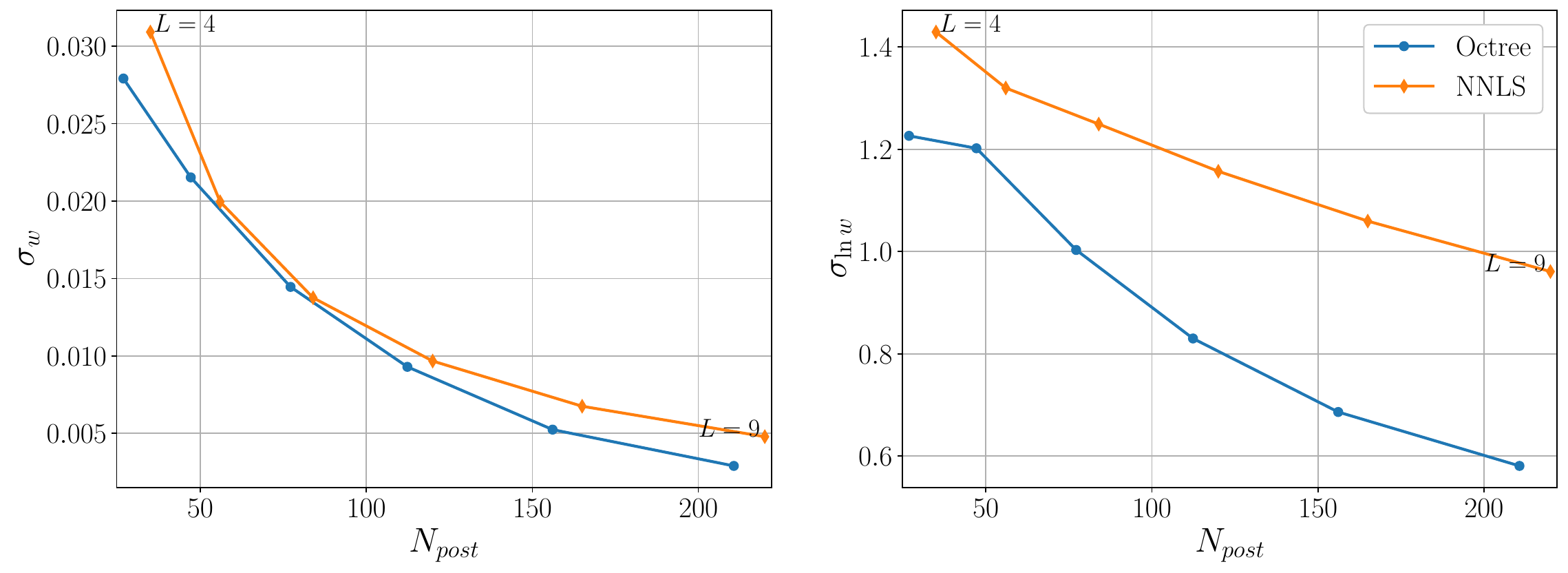}
    \caption{Standard deviation of the post-merge particle weights (left) and logarithms of the weights (right), pre-merge particles with equal weights.}\label{fig:merging-sigma-equalweight}
    \end{figure}
Looking at the distribution of post-merge particle weights, as shown on Figure~\ref{fig:merging-sigma-equalweight}, we see that although the overall variances of the weights are similar, the NNLS merging
produces more particles with very small weights, as evidenced by the larger values of the standard deviation of the logarithms of the post-merge particles' weights.

  \begin{figure}[t]
    \centering
    \includegraphics[width=0.49\textwidth]{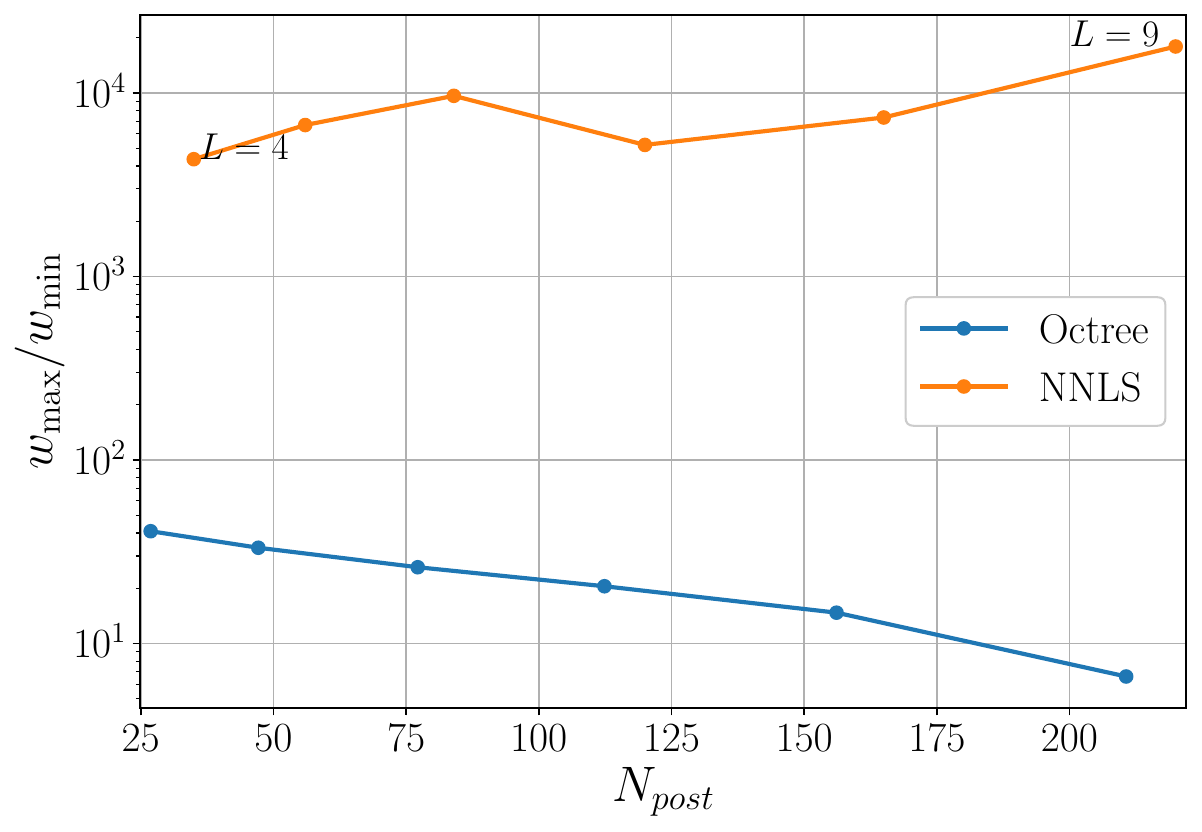}
    \caption{Ratio of largest to smallest post-merge weight, pre-merge particles with equal weights.}\label{fig:merging-ratio-equalweight}
    \end{figure}
This is corroborated by the results shown on Figure~\ref{fig:merging-ratio-equalweight}, where it can be seen that NNLS produces a maximum weight discrepancy two orders of magnitude higher than that
produced by the octree merging routine. Moreover, whilst refinement of the octree bins (i.e. a larger number of post-merge particles) leads to a more uniform distribution of particle weights, NNLS still
retains very low-weight particles, as it can be seen on Figure~\ref{fig:merging-ratio-equalweight} that the ratio does not decrease even as the overall variance (as seen on Figure~\ref{fig:merging-sigma-equalweight})
decreases. The larger degree of weight non-uniformity exhibited by the NNLS merging in this scenario might lead to higher
costs of evaluating collisions in a full simulation~\cite{schmidt2000new}.

\subsubsection{Weighted samples}
Now, we consider an alternative sampling strategy. We sample each particle's velocity components independently from a uniform distribution on the interval $[-4 v_{\mathrm{ref}}, 4 v_{\mathrm{ref}}]$,
where $v_{\mathrm{ref}} = \sqrt{2kT/m}$. The particles' weights are then computed as
\begin{equation}
  w_i = \frac{1}{C_w}\exp \left(-\frac{m ||v||_2^2}{2kT} \right),
\end{equation}
where $C_w$ is a scaling constant to ensure the correct total number density. In this case, the pre-merge particle weights already exhibit very strong variations in their magnitudes.

\begin{figure}[t]
  \centering
  \includegraphics[width=0.98\textwidth]{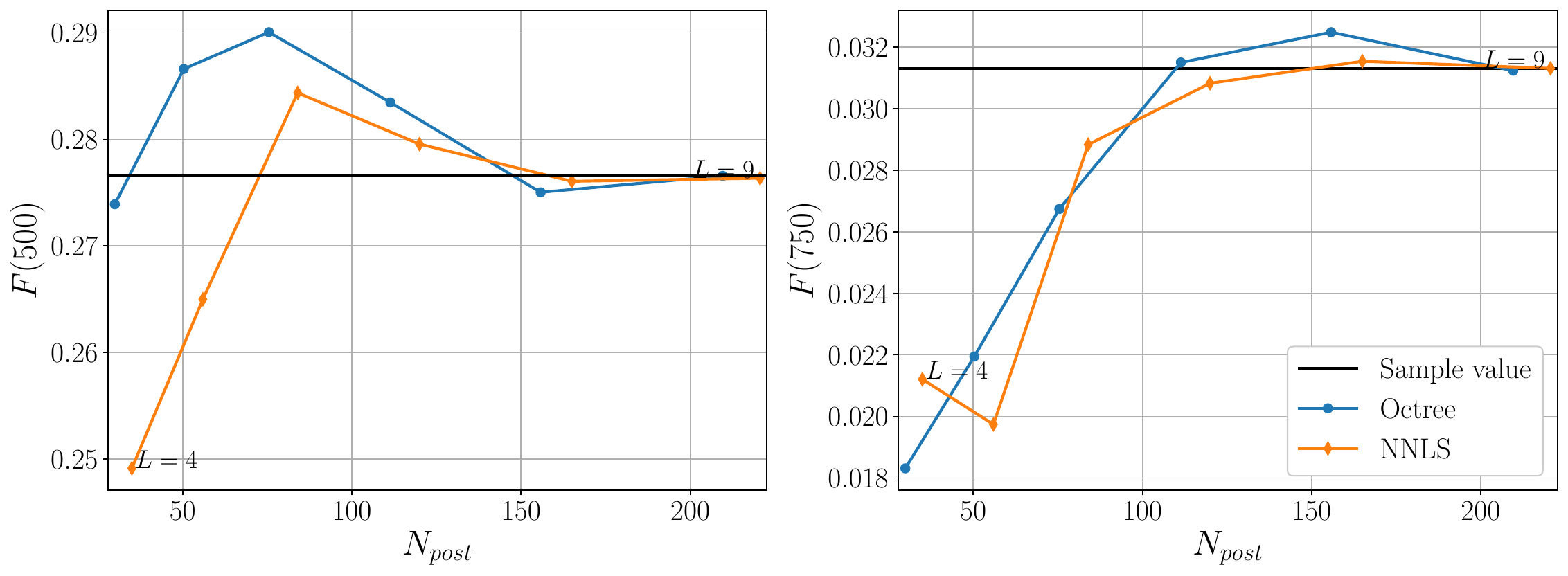}
  \caption{Values of the tail functions $F(500)$ (left) and $F(750)$ (right) as a number of post-merge particles, pre-merge particles with unequal weights.}\label{fig:merging-tf-velbox}
  \end{figure}
  Figure~\ref{fig:merging-tf-velbox} shows the values of the tail functions $F(500)$ and $F(750)$ as a number of post-merge particles, with the initial value of the ensemble shown in black.
For this sampling strategy, both algorithms show similar behvaiour in terms of accuracy of the representation of the tails of the velocity distribution. It should however be noted that the scale on
the right subplot (showing $F(750)$) differs from that shown on Figure~\ref{fig:merging-tf-equalweight}, as the octree merging has noticeably lower error in the case of uneqally weighted particles.

  \begin{figure}[t]
    \centering
    \includegraphics[width=0.98\textwidth]{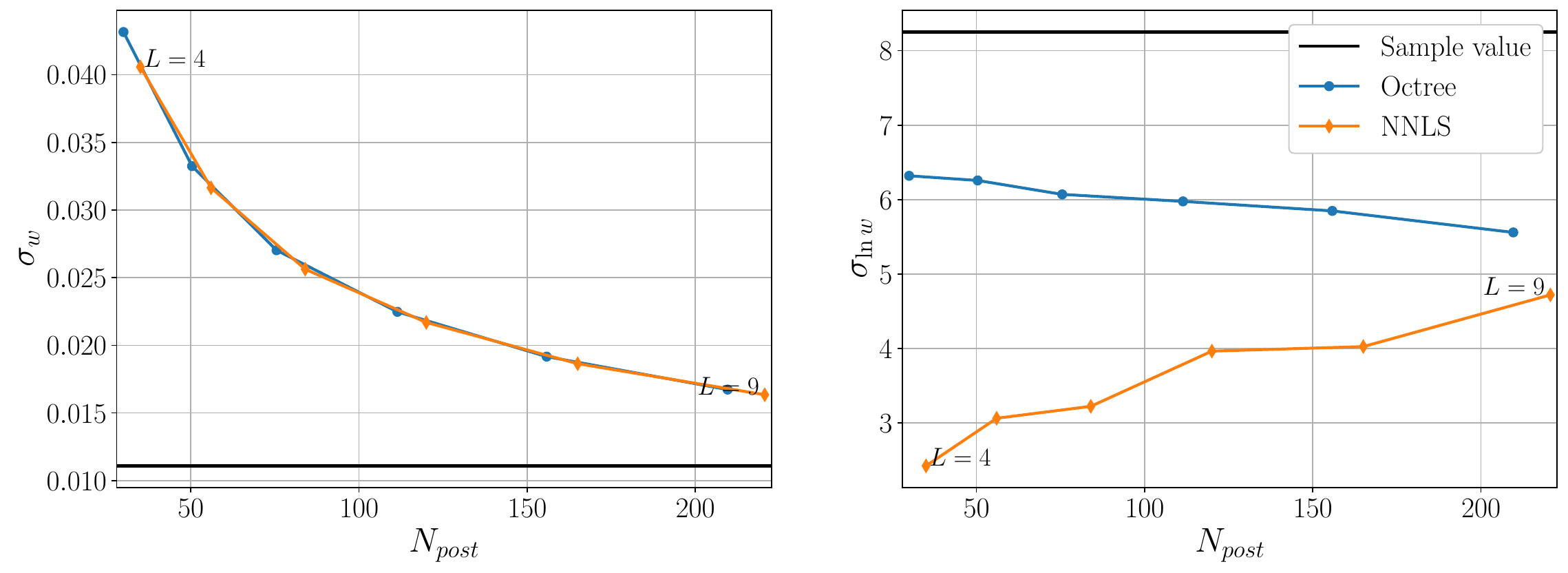}
    \caption{Standard deviation of the post-merge particle weights (left) and logarithms of the weights (right), pre-merge particles with unequal weights.}\label{fig:merging-sigma-velbox}
    \end{figure}
    Figure~\ref{fig:merging-sigma-velbox} shows the post-merge variances in the particle weights, along with the variance of the weights of the initial sample.
    Again, the overall variances of the weights are similar, but in this case, the NNLS merging produces noticeably fewer particles with very small weights, as seen on the right subplot of the figure.

  \begin{figure}[t]
    \centering
    \includegraphics[width=0.49\textwidth]{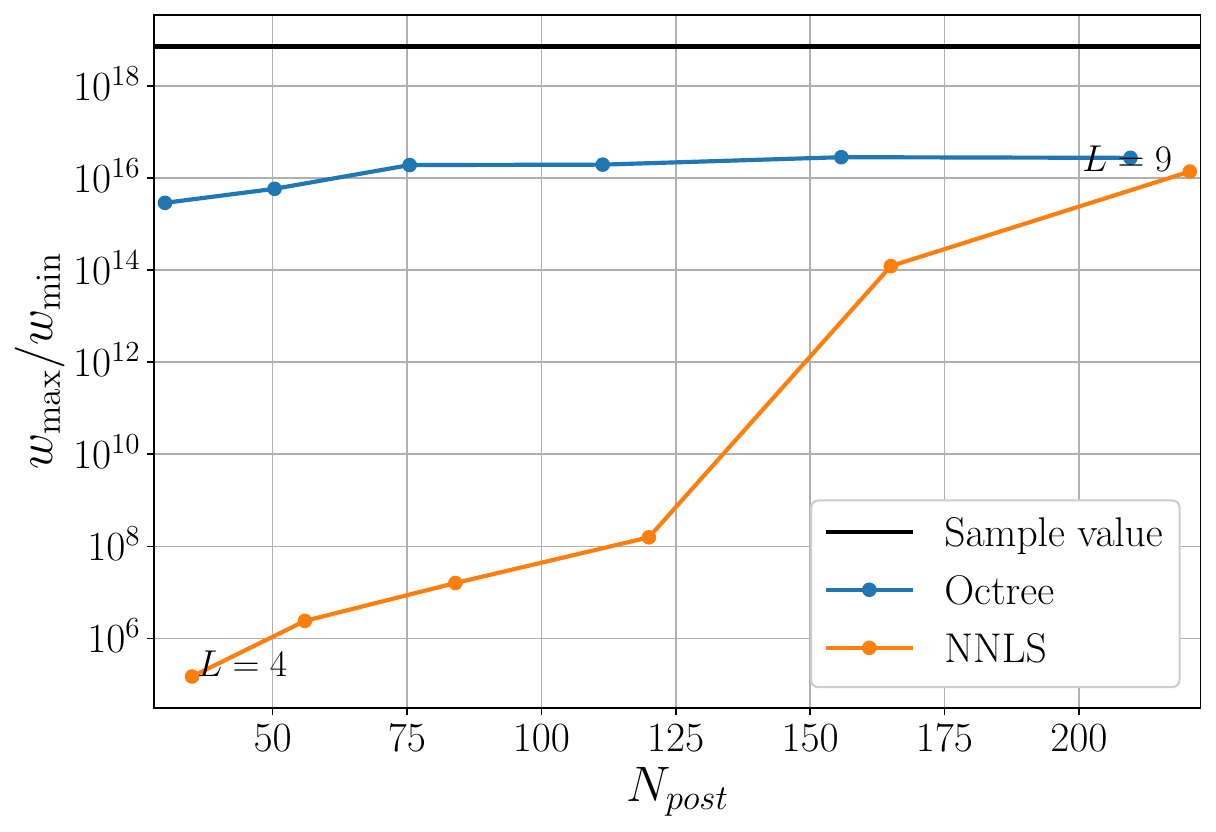}
    \caption{Ratio of largest to smallest post-merge particle weight, pre-merge particles with unequal weights.}\label{fig:merging-ratio-velbox}
    \end{figure}
The ratio of largest to smallest post-merge weight, as shown on Fig.~\ref{fig:merging-ratio-velbox}, is also significantly (up to 8 orders of magnitude) lower for the NNLS merging approach.
Concluding, we can summarize that regardless of the sampling strategy deployed, the NNLS merging routine preserves the high-speed tails of the distribution function with more accuracy
than an octree-based particle grouping with group-wise merging. In case of a pre-merge distribution with equal particle weights, NNLS merging leads to larger discrepancies in the post-merge particle weights,
which potentially has an impact on the cost of evaluating collisions~\cite{schmidt2000new}. However, if the pre-merge distribution has unequal particle weights, NNLS merging exhibits a noticeably smaller
discrepancy in the weights compared to the octree merging algorithm.

\subsection{0D simulations}
\subsubsection{BKW relaxation}
Next, we consider the Bobylev-Krook-Wu (BKW) relaxation problem~\cite{bobylev1976,krook1977}, an analytical solution to the Boltzmann equation for pseudo-Maxwell molecules.

We initialize the particle velocities on a spherically cut-off grid in velocity space with from $-v_{max}$ to $v_{max}$ in each direction:
\begin{equation}
  \mathbf{v} \in \left\{\left(v_{x,i},v_{y,j},v_{z,k}\right) | v_{x,i}^2+v_{y,j}^2+v_{z,k}^2 \leq v_{max}^2  \right\},
\end{equation}
where
\begin{equation}
  v_{r,i} = -v_{max} + i \Delta v,\:r=x,y,z,\:i=0,N_g,\quad \Delta v = \frac{2v_{max}}{N_g}.
\end{equation}
Here $N_g$ is the number of grid points, chosen as 36 in the present study, and $v_{max}$ was taken to be equal to $4\sqrt{\frac{2kT}{m}}$, leading to 22400 particles being initialized.
The particles' weights are computed by evaluating the BKW distribution at $t=0$ at the velocity grid nodes and then re-scaling to obtain the prescribed number density $n$:
\begin{equation}
  w(\mathbf{v}) \propto  ||\mathbf{v}||_2^2 \exp\left(-\frac{5  ||\mathbf{v}||_2^2 m }{6 kT}\right).
\end{equation}

We define the ``total moment'' of order $2l$ as 
\begin{equation}
  {M}_{2l} = \frac{1}{n} \int ||\mathbf{v}||_2^{2l} f(\mathbf{v}) \mathrm{d} \mathbf{v},
\end{equation}
and the corresponding scaled total moment, scaled with respect to the corresponding total moment of the Maxwell--Boltzmann distribution ${M}_{2l}^{M-B}$ at the same temperature $T$:
\begin{equation}
  \hat{M}_{2l} = \frac{{M}_{2l}}{{M}_{2l}^{M-B}},\: {M}_{2l}^{M-B} = 2 \pi \left( \frac{m}{2\pi kT} \right)^{3/2} \Gamma\left(\frac{3+2l}{2}\right)\left(\frac{2kT}{m}\right)^{\frac{3+2l}{2}}.
\end{equation}
Finally, we introduce the scaled time as 
\begin{equation}
  \hat{t} = \frac{1}{n \pi d^2} \sqrt{m}{2\pi kT}.
\end{equation}
The analytical expression for the time evolution of the scaled moments of the BKW distribution then reads
\begin{equation}
  \hat{M}_{2l}^{an}(\hat{t}) = C^{l-1} \left(l - (l-1) C\right),\: C = C(\hat{t}) = 1 - \frac{2}{5}\exp\left(-\hat{t} \frac{2\sqrt{2}}{6 \sqrt{\pi}}\right).
\end{equation}

We choose argon ($m=66.3 \cdot 10^{-27}$ kg) as the working gas, a temperature $T=237$~K, a number density $n=10^{23}$~m$^{-3}$, and a molecular diameter $d=4.11$ \AA. The simulations are performed with a time-step
of $0.025 \hat{t}$ for 600 timesteps, and the variable-weight No Time Counter scheme proposed in~\cite{schmidt2000new} is used for the evaluation of collisions. Since the particles have unequal weights at the start of the simulation, even elastic collisions lead to
particle splitting, thus necessitating merging to keep particle numbers within a certain range.
We choose $L \in \left\{4,5,6,7,8,9 \right\}$,
and perform merging when the number of particles exceeds approximately 120~\% of the number of expected post-merge
particles (i.e. the total number of moments being conserved). We also run simulations with the octree merging algorithm~\cite{martin2016octree}, where we choose the same threshold and post-merge numbers of particles
as in the NNLS merging algorithm. We perform ensemble averaging to ensure a sufficiently low level of stochastic noise, with averaging over 10800 ensembles for $L=4$, 4800 for $L=5$, 2160 for $L=6$,
1040 for $L=7$, 560 for $L=8$, 400 for $L=9$, and the same number of ensembles used for the corresponding simulations with the octree merging.

\begin{figure}[t]
\centering
\includegraphics[width=0.98\textwidth]{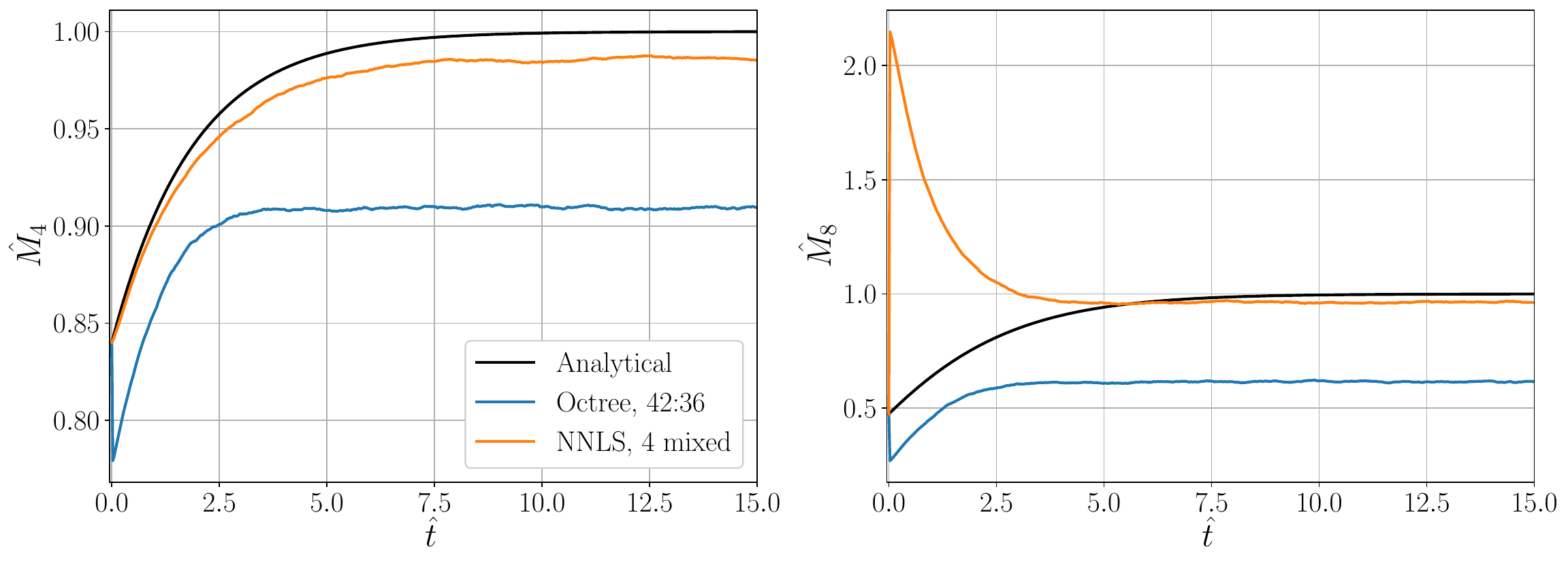}
\caption{Evolution of the 4$^{th}$ (left) and 8$^{th}$ (right) scaled moments for $L=4$.}\label{fig:bkw-nnlsupto4}
\end{figure}

Figure~\ref{fig:bkw-nnlsupto4} shows the evolution of 4$^{th}$ (left) and 8$^{th}$ scaled total moments for $L=4$, which corresponds to merging being performed when the particle count exceeds 42 particles,
and the post-merge number of particles is approximately 36.
Since the simulation is initialized with a very large number of particles, merging is performed immediately at the first timestep, and this leads to a significant jump in the value of the computed moment
in case it is not conserved by the merging algorithm. Since conservation of all mixed moments up to order 4 is sufficient to conserve $\hat{M}_{4}$, the simulation using the NNLS merging exhibits no such jump,
and in general its values are closer to that of the analytical solution, compared to the simulation results obtained with the octree merging algorithm. For the total moment of order 8, the NNLS merging-based
results also display a significant deviation from the computed value of the moment at the start of the simulation; however, once the system reaches steady state, the values of the moment are still closer to those
of the analytical solution when compared to the results of the octree merging. This can be attributed to the fact that even though moments of order 5 and higher are not conserved, conservation of various lower-order
moments still preserves some of the collisional dynamics of the system.

\begin{figure}[t]
  \centering
  \includegraphics[width=0.98\textwidth]{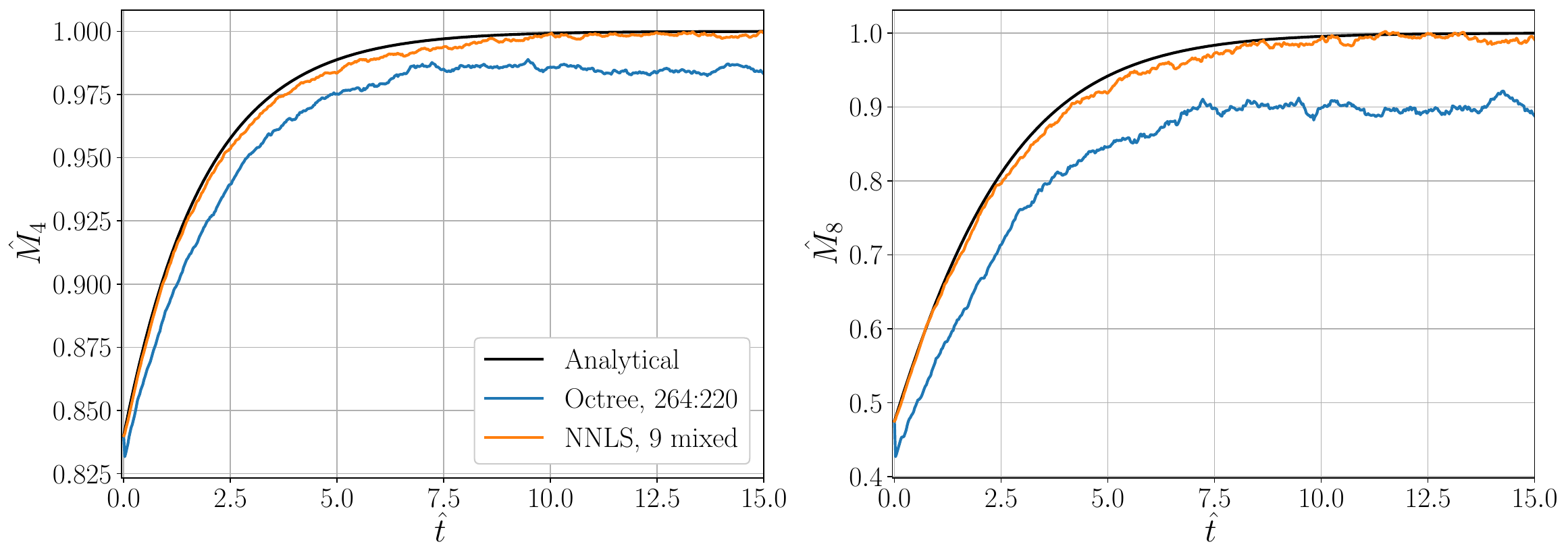}
  \caption{Evolution of the 4$^{th}$ (left) and 8$^{th}$ (right) scaled moments for $L=9$.}\label{fig:bkw-nnlsupto9}
  \end{figure}

The case of conservation of all mixed moments up to total order 9 is shown on Figure~\ref{fig:bkw-nnlsupto9}. Merging was performed when the particle count exceeded 264 particles, and the post-merge number of particles was approximately 220. Since enough moments are preserved in the NNLS merging to ensure conservation of both the 4$^{th}$ (left) and 8$^{th}$ total moments, the simulations do not exhibit any
jumps in the values at the start, and lie very close to the analytical solution.

To better compare the two merging approaches, we consider the bias of the ensemble-averaged solution w.r.t. the analytical solution:
\begin{equation}
    \mathcal{B}\left(\hat{M}_{2l}\right) =\sqrt{ \frac{1}{N_t}  \sum_{\hat{t}_i} \left(\overline{\hat{M}}_{2l}(\hat{t}_i) - \hat{M}_{2l}^{an}(\hat{t}_i)\right)^2 }.
\end{equation}
Here $N_t$ is the number of timesteps for which the bias is computed, $\hat{t}_i = i\Delta t$ is the time at timestep $i$, $\overline{\hat{M}}_{2l}$ is the ensemble average of the moments computed with the given set of simulation parameters, and $\hat{M}_{2l}^{an}$ is the analytical value of the moment.
\begin{figure}[t]
  \centering
  \includegraphics[width=0.98\textwidth]{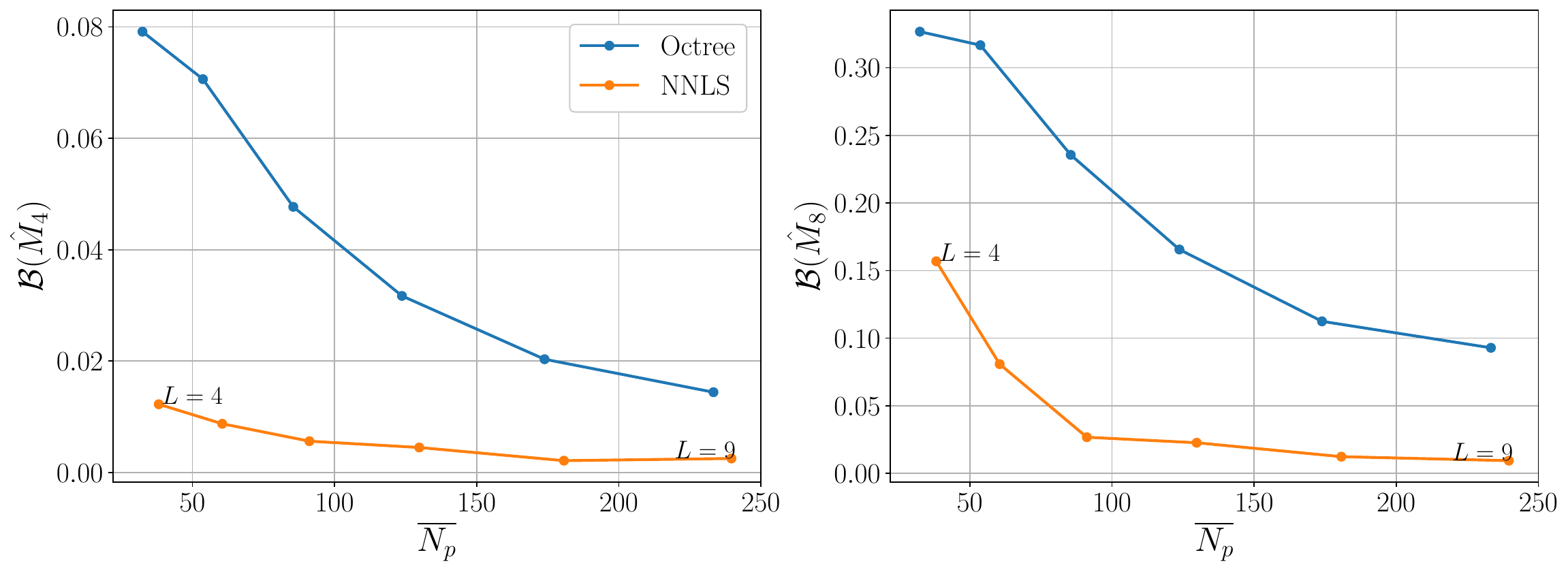}
  \caption{Bias in the 4$^{th}$ (left) and 8$^{th}$ (right) scaled moments as a function of the average number of particles in the simulation.}\label{fig:bkw-bias}
  \end{figure}

Figure~\ref{fig:bkw-bias} shows the bias in the 4$^{th}$ and 8$^{th}$ scaled moments as a function of the time- and ensemble-averaged number of particles in the simulation $\overline{N}_p$. It can be observed
that use of the NNLS merging leads to significantly lower bias in the computed values of the moments, especially when a relatively small number of particles is used. \blue{As the total order of conserved moments $L$ is increased, the bias in the computed moments is reduced, as expected, and remains noticeably lower than that obtained with the octree merging approach for all of the average numbers of particles considered in the simulation.}

\begin{figure}[t]
  \centering
  \includegraphics[width=0.98\textwidth]{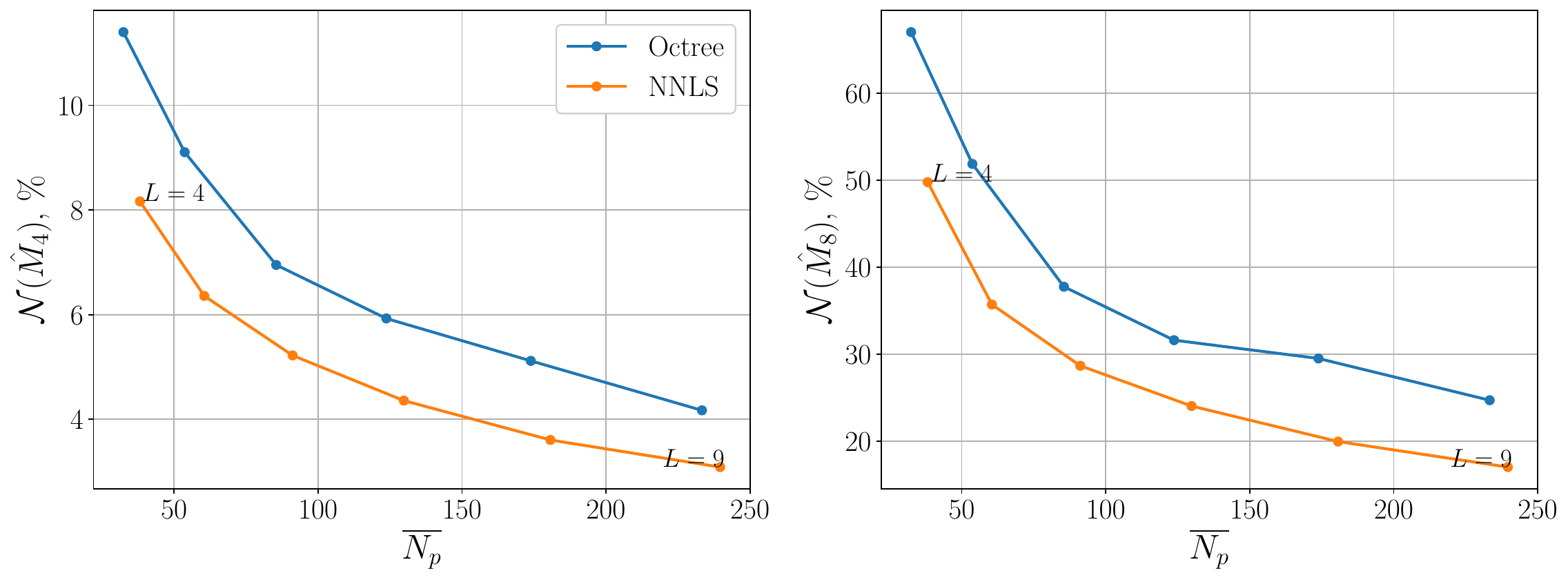}
  \caption{\blue{Relative noise in the 4$^{th}$ (left) and 8$^{th}$ (right) scaled moments as a function of the average number of particles in the simulation.}}\label{fig:bkw-noise}
  \end{figure}

\blue{Finally, we investigate the relative noise in the computed moments as a function of the average number of particles in the simulation. Figure~\ref{fig:bkw-noise} shows the time-averaged relative (with respect
to the ensemble average) noise in the 4$^{th}$ and 8$^{th}$ scaled moments as a function of the average number of particles in the simulation. We see that use of the NNLS merging reduces the level of stochastic noise in the computation compared to the octree-based
approach; however, the overall scaling with the number of particles is similar in both cases.}

\subsubsection{Ionization with constant electric field}\label{sec:ion}
We next consider the case of a spatially homogeneous Ar/Ar$^+$/e$^-$ plasma accelerated by a constant electric field.
We consider elastic and ionizing electron-neutral collisions only, and performs collisions according to the scheme as described in~\cite{nanbu2002probability}, with
the post-collision energy in case of an ionizing collision divided equally between the primary and secondary electrons.
For simplicity, we assume isotropic scattering for all interactions, and use the IST Lisbon~\cite{alves2014lisbon} cross-section data for
the electron-neutral interactions.

For a given value of the reduced electric field, the plasma reaches a quasi-steady state characterized by constant collision rate coefficients
and a constant electron temperature; after a while, all the neutrals are depleted and the electron-neutral collision rates become zero.
It is thus possible to extract the ionization rate coefficients and electron temperature values from the unsteady simulation data (omitting the
initial transient and later timesteps where the neutral population becomes very low) and study the impact of particle merging on these quantities. The initial number densities of the ions and electrons
were taken to be $10^{-7} n_{n}$, where the number density of the neutrals $n_n$ was taken as $10^{23}$ m$^{-3}$. The temperature of the neutrals was taken as 300~K, and that of the electrons as 2~eV.

The event splitting scheme~\cite{oblapenko2022hedging} is used for the collisions for two reasons:
\begin{enumerate}
  \item it provides a significant reduction of stochastic noise in the values of the ionization rate coefficient
  \item it leads to a faster growth rate of the number of simulation of particles and as an effect, more frequent particle merging, amplifying the effects of the merging procedures.
\end{enumerate}
We consider two values of the reduced applied electric field: 100 and 400~Tn.
To verify the implementation of the electron-neutral collisions, we first run simulations using the octree merging with the target number of electron particles set to 6000 and the threshold number of electron particles set to 12000. A timestep of 50~fs is used, and averaging is performed over 3500000 and 350000 timesteps for the field strengths of 100 and 400 Tn, respectively. We compare the computed values of the ionization rate coefficient to those obtained with the use of the
Bolsig+ solver~\cite{hagelaar2005solving}.

\begin{table}[h!]
  \centering
\begin{tabular}{ c || c | c }
 Field strength, Tn & Bolsig+,  & Present work \\ 
 \hline
 100 & $3.771 \cdot 10^{-16}$ & $3.768 \cdot 10^{-16}$ \\  
 400 & $4.382 \cdot 10^{-15}$ & $4.462 \cdot 10^{-15}$    
\end{tabular}
\caption{All values in m$^3 \cdot$ s$^{-1}$.}
\label{tab:kion}
\end{table}
The computed values of the ionization rate coefficient are given in Table~\ref{tab:kion}. For the lower reduced field strength, the discrepancy
between the solutions is less than 0.1\%, but increases up to almost 2\% for the reduced field strength of 400~Tn.
This can partially be attributed to larger errors of the two-term approximation used in the Bolsig+ solver at higher field strengths,
slight differences in the collisions mechanics being modelled, as well as differences in how the tabulated cross-sections are interpolated.
Nevertheless, we assume that the errors are sufficiently small as to show that the variable-weight DSMC code used in the present work
is capable of modelling ionization phenomena with a sufficient degree of accuracy. We use the values computed using the DSMC code as given in the table above as reference values against which we
compare results obtained with smaller number of particles and more frequent merging.

We vary the following parameters for the different merging algorithms:
\begin{itemize}
  \item The target post-merge particle numbers for the electrons in the case of octree merging, ranging from 38 to 220.
  \item The number of preserved mixed velocity moments $L$ in the case of NNLS merging, ranging from 4 to 9.
  \item In the case of NNLS merging, we perform merging without rate preservation (``NNLS''), with exact rate preservation (``NNLS RP''), and with approximate rate preservation (``NNLS, ARP'').
\end{itemize}

\begin{figure}[t]
  \centering
  \includegraphics[width=0.98\textwidth]{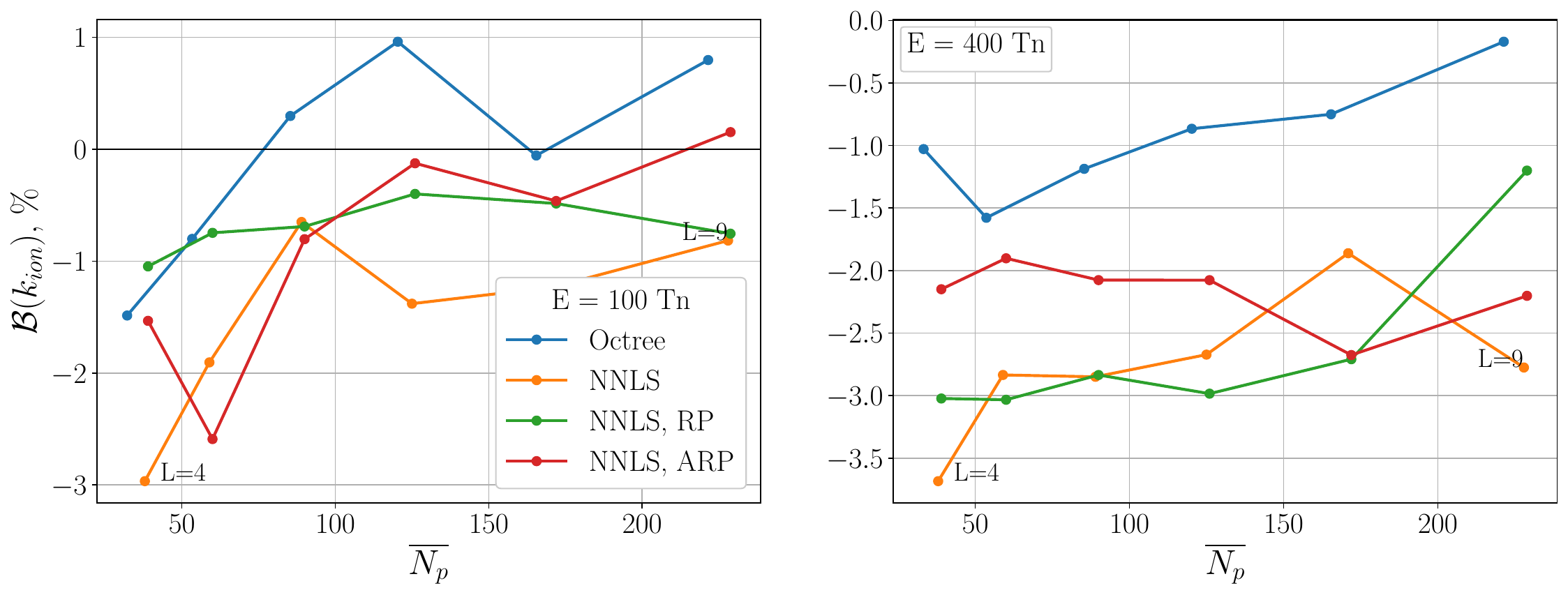}
  \caption{Average bias in the ionization rate coefficient as a function of the average number of particles in the simulation for an applied electric field of 100~Tn (left) and 400~Tn (right).}\label{fig:kion-bias}
  \end{figure}

\begin{figure}[t!]
  \centering
  \includegraphics[width=0.98\textwidth]{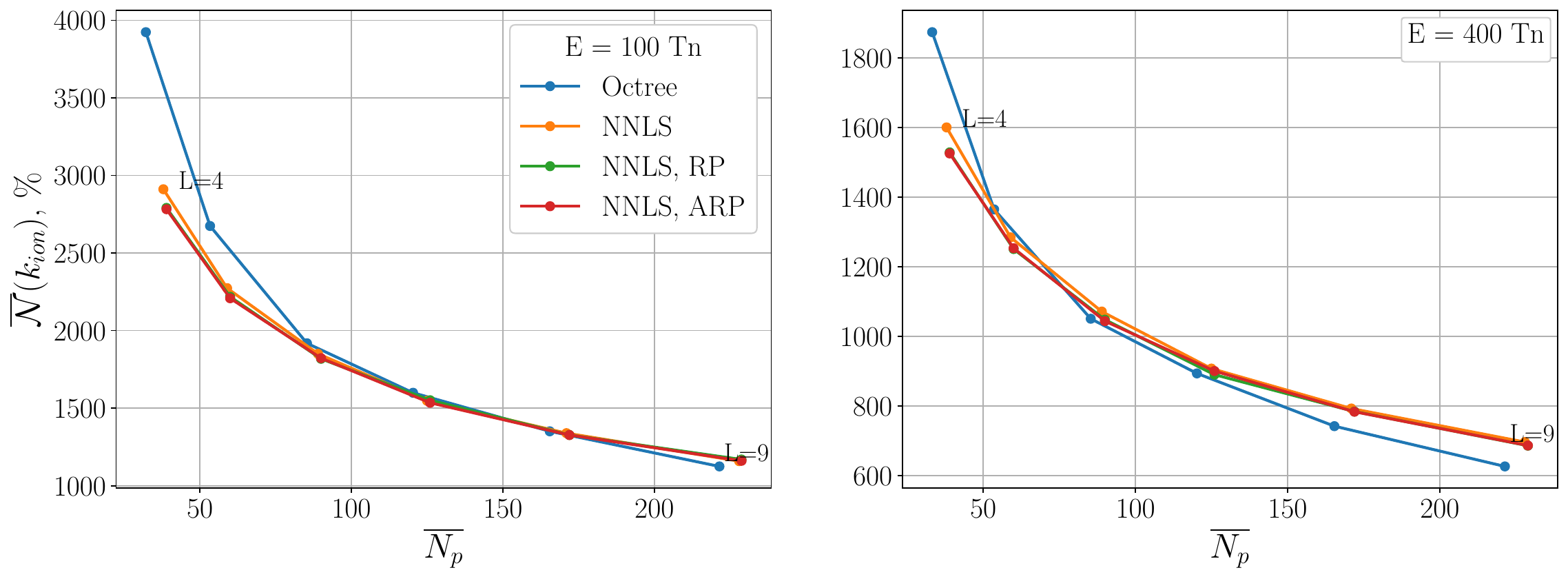}
  \caption{Average noise in the ionization rate coefficient as a function of the average number of particles in the simulation for an applied electric field of 100~Tn (left) and 400~Tn (right).}\label{fig:kion-noise}
  \end{figure}

  \begin{figure}[t!]
    \centering
    \includegraphics[width=0.98\textwidth]{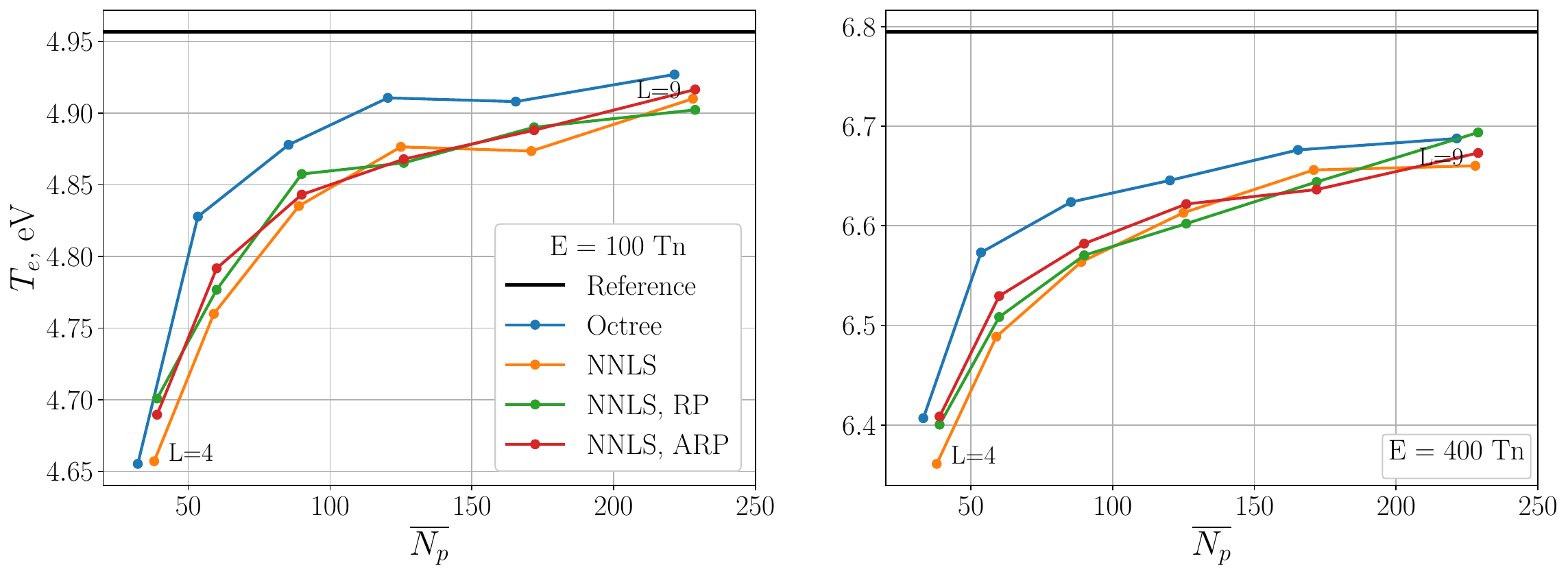}
    \caption{Computed electron temperature as a function of the average number of particles for an applied electric field of 100~Tn (left) and 400~Tn (right).}\label{fig:Te}
    \end{figure}

The threshold number of particles that defines
when merging is carried out for the electrons is set at approximately 108\% of the the target number of particles.
To reduce the impact of stochastic noise, perform ensemble averaging: for each set of parameters with the merging threshold number 
of particles less than 100, we run 64 simulations; for higher merging threshold values, 16 independent ensembles are averaged.

Figure~\ref{fig:kion-bias} shows the bias in the computed electron-impact ionization rate coefficient. We see that the bias is quite low for all merging approaches, however, the octree-based merging
performs slightly better. Preservation of electron-neutral collision rates, whether exact (``NNLS, RP'') or approximate (``NNLS, ARP'') has a minor impact at the lowest particle counts, but does not affect the results significantly.

  Next, we consider the stochastic noise in the ionization rate coefficient, as shown on Figure~\ref{fig:kion-noise}. Use of NNLS merging leads to a lower level of noise at low particle count compared to the
  octree merging, with incorporation
  of approximate rate preservation lowering the noise somewhat further, albeit by a very small amount. Both the approximate and exact-rate preserving approaches produce virtually indistinguishable results. At higher particle counts, both the NNLS and octree merging approaches lead to very similar results.

   Finally, we look at the computed electron temperature, as shown on Figure~\ref{fig:Te}. For this quantity, use of the octree-merging leads to values closest to the reference values. Use of
   approximate rate preservation in the NNLS merging again slightly reduces the error at lower particle counts. However, preservation of exact rates (``NNLS, RP'') does lead to an improvement in the electron temperature error, especially for the higher field strengths.
    \blue{This is in line with the results shown on Figure~\ref{fig:merging-tf-velbox}, where the lower-velocity part of the distribution exhibits a higher level of distortion when NNLS merging is used.}

\begin{figure}[t!]
  \centering
  \includegraphics[width=0.98\textwidth]{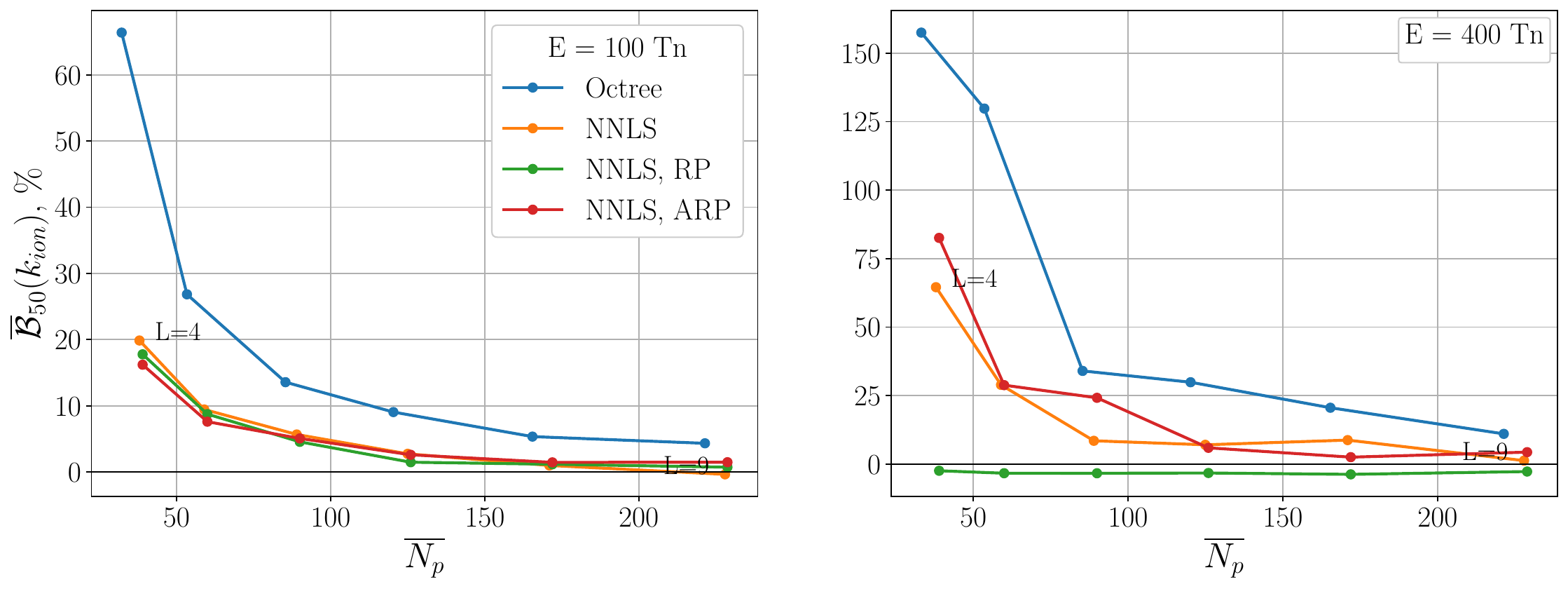}
  \caption{Average bias in the ionization rate coefficient as a function of the average number of particles in the simulation for an applied electric field of 100~Tn (left) and 400~Tn (right) in 
  the first 50 timesteps after merging.}\label{fig:kion-bias-50}
  \end{figure}

\begin{figure}[t!]
  \centering
  \includegraphics[width=0.98\textwidth]{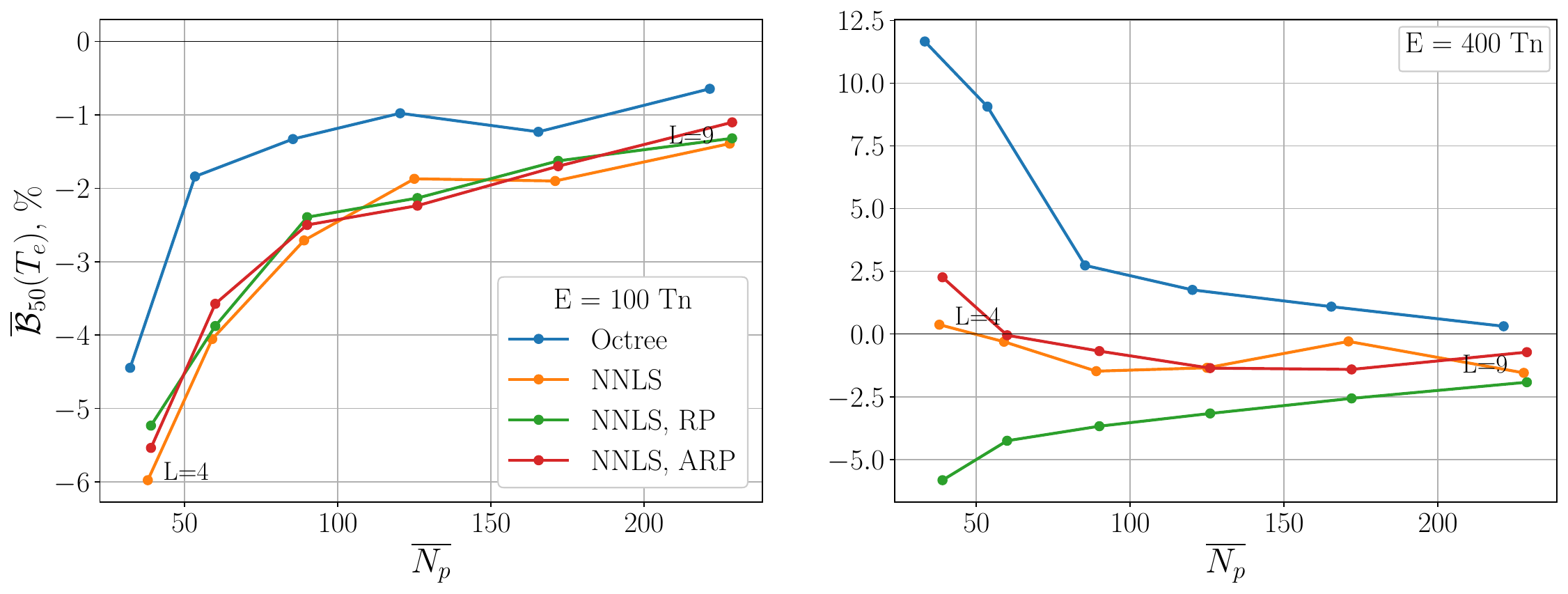}
  \caption{Average bias in the electron temperature as a function of the average number of particles in the simulation for an applied electric field of 100~Tn (left) and 400~Tn (right) in 
  the first 50 timesteps after merging.}\label{fig:Te-bias-50}
  \end{figure}

Until this point, we considered properties of the plasma system averaged over the whole quasi-steady state timeframe. However, the average time between merging events in the simulations is quite large, especially for lower particle counts. This means the whole particle system has time to ``recover'' from the
error induced by the merging, as there is a constant driving force present (the constant electric field). For a coupled PIC simulation the effect of the merging could be more noticeable, as the induced error could affect the
evolution of the electric field, thereby affecting the further behaviour of the particles.
Therefore, we also look at the bias in the average ionization rate coefficient and electron temperature in the first 50 timesteps after a merging event, averaged over all merging events.

  Figures~\ref{fig:kion-bias-50}--\ref{fig:Te-bias-50} show the average bias in the computed ionization rate coefficient and electron temperature in the first 50 timesteps after a particle merging event,
  respectively.
  We see that for the ionization rate coefficient, especially at lower field values, the moment-preserving NNLS merging significantly outperforms the octree merging; \blue{and for the higher field strength,
  the exact-rate preserving approach outperforms the other ones.}
  Approximate preservation of the ionization rate coefficient has little impact on the results; and as more higher-order moments are preserved, the benefits of conserving the approximate rates
  become even less noticeable. This is expected, as the ionization is driven by the high-speed electrons, therefore, a merging procedure that ensure accurate representation of the tails of
  the distribution function (e.g. by conserving high-order moments) will also ensure an accurate representation of processes affected by the tails of the distribution function.
  For the temperature, similar to the previous results, the octree merging exhibits less bias immediately after the merging compared to the NNLS approach. Rate preservation again has a very limited
  impact at low particle numbers, with almost no difference between the approximate- and exact rate-preserving schemes.
  
  \blue{We see that the ionization rate bias immediately after merging is much lower for the NNLS-based approaches, but this is not the case when looking at the average over the whole qausi-steady part of the simulation. We assume that this under-performance of NNLS compared to the octree merging is due to the fact that NNLS basically chooses a new support for the distribution function during each merging call,
  thus leading to strong jumps in the distribution. Although the conservation of high-order moments (and potentially collision rates) ensures a lower bias immediately after merging,
  the distortion in the distribution function over successive merges leads to a higher bias in the long run. To test this hypothesis, we devise the following test case: similarly to the test case presented in~Sec.~\ref{sec:equal-weight}, we sample 500 particles either from a fixed-weight or a variable-weight distribution, and merge them via the NNLS or octree merging approaches. We then run the same merging procedure on an identical set of particles, but which have had a small velocity perturbation applied to them, sampled from a normal distribution with a standard deviation $\sigma_v$ of $10^{-3}$ or $10^{-2}$ and then multiplied by the thermal velocity. We then compute the weighted chamfer distance $d_c$ between the two-post merge particle sets. We repeat this procedure 2000 times with different random seeds, and then compute the RMS $\overline{d}_c$ of the chamfer distance over the ensembles. We then analyze the ratio of $\overline{d}_c$ to the average pre-merge displacement of the particles, given by $\sqrt{3}\sigma_v$.}
\begin{figure}[t!]
  \centering
  \includegraphics[width=0.98\textwidth]{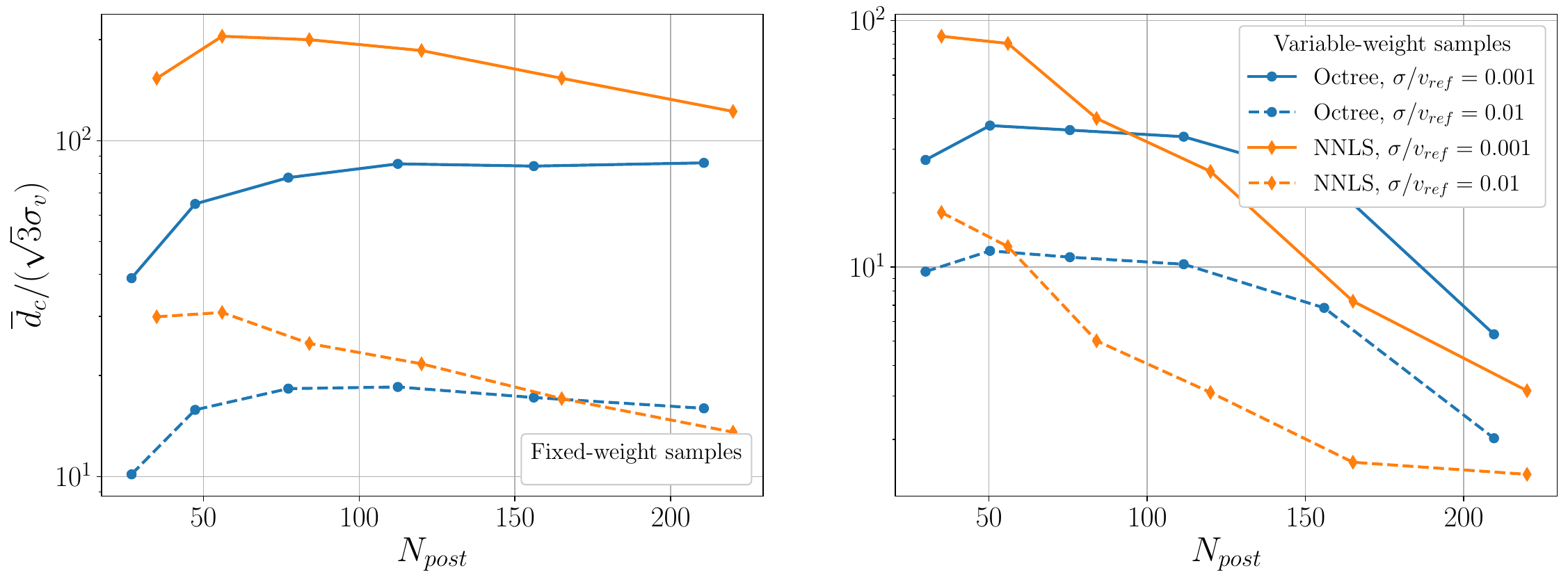}
  \caption{\blue{Ratio of relative chamfer distance between post-merge distributions for non-perturbed and perturbed particles as a function of the number of post-merge particles. Equal
  pre-merge particle weights (left), unequal pre-merge particle weights (right).}}\label{fig:chamfer-dist}
  \end{figure}
  \blue{We see that the NNLS merging is more strongly affected by minor variations in the distribution,
        as it effectively chooses a new lower-dimensional support at each merging step, whereas the octree bins' bounds vary smoothly with the particles'
        velocities (although the choice of the next bin to refine varies non-smoothly). This also explains why the rate-preserving approaches work well only when one considers the bias immediately after the merging event - by construction, they preserved approximate/exact ionization rates. Over 50 timesteps, the energy gained by an electron is approximately 10$^{-3}$ eV in the case of the 400~Tn field, that is, any significant distortion in the distribution function, especially in the lower-energy part (where NNLS is more sensitive to the number of moments chosen) does not propagate into the high-energy region above the ionization threshold of argon. Over larger timescales,
        however, the stronger susceptibility of NNLS to the distribution function's perturbations leads to a larger bias in the ionization rates, which also affects the electron energy temperature.}
  
  \blue{Therefore, future work will also explore the effect of coupling of the NNLS merging with an octree velocity space binning in order to both reduce the computational cost of NNLS and to
  limit the changes in the distribution function introduced by the merging.}
  Future work will explore the impact of exact and approximate rate
  preservation in the case of hot neutrals,
  where their velocities are non-negligible.

\subsection{1D simulation of Fourier flow}\label{sec:fourier}
Finally, we consider a temperature gradient-induced flow between two plates. The flow is simulated in a 1-dimensional setup. The distance between the plates is taken as $L=5$~mm, the temperature
of the left wall is taken to be 300~K, and the temperature of the right wall is taken to be 600~K. The number density is taken as $3.14526 \times 10^{22}$ m$^{-3}$, corresponding to a Knudsen
number of approximately $10^{-2}$. 1000 uniform cells are used for the domain discretization, each cell thus having a size of roughly 1/10$^{th}$ of the mean free path.
A timestep of 40~ns is used. Argon with the VHS
cross-section model is used as the test gas. We perform averaging of the macroscopic flow quantities over $3\times10^{6}$ timesteps after the first 500000 timesteps have passed to ensure that
the flow has reached steady state. As a reference solution, we consider a fixed-weight simulation with a 500000 particles.

For both variable-weight simulations, 500 equal-weight particles are sampled in each cell and immediately merge down to the target number
of particles. Subsequent merging is performed when the number of particles in a cell exceeds 120\% of the target number of particles;
for NNLS merging, the target number of particles is set similarly to the BKW case: we consider conservation of all mixed velocity moments up to order $L$, and we choose $L \in \left\{4,5,6,7,8\right\}$; in addition, we also conserve the 1-st and 2-nd spatial moments of
the particle distribution in the $x$-direction. For octree merging, the target number of particles was chosen to be the same as
the number of post-merge particles in the NNLS approach; the $N:2$ merging performed in each octree bin also conserves the 1-st
and 2-nd spatial moments of the particle distribution~\cite{martin2016octree}, unless post-merge particles' positions end outside
of the domain, in which case they are simply set to be inside the domain at the domain boundary.

\begin{figure}[t]
  \centering
  \includegraphics[width=0.98\textwidth]{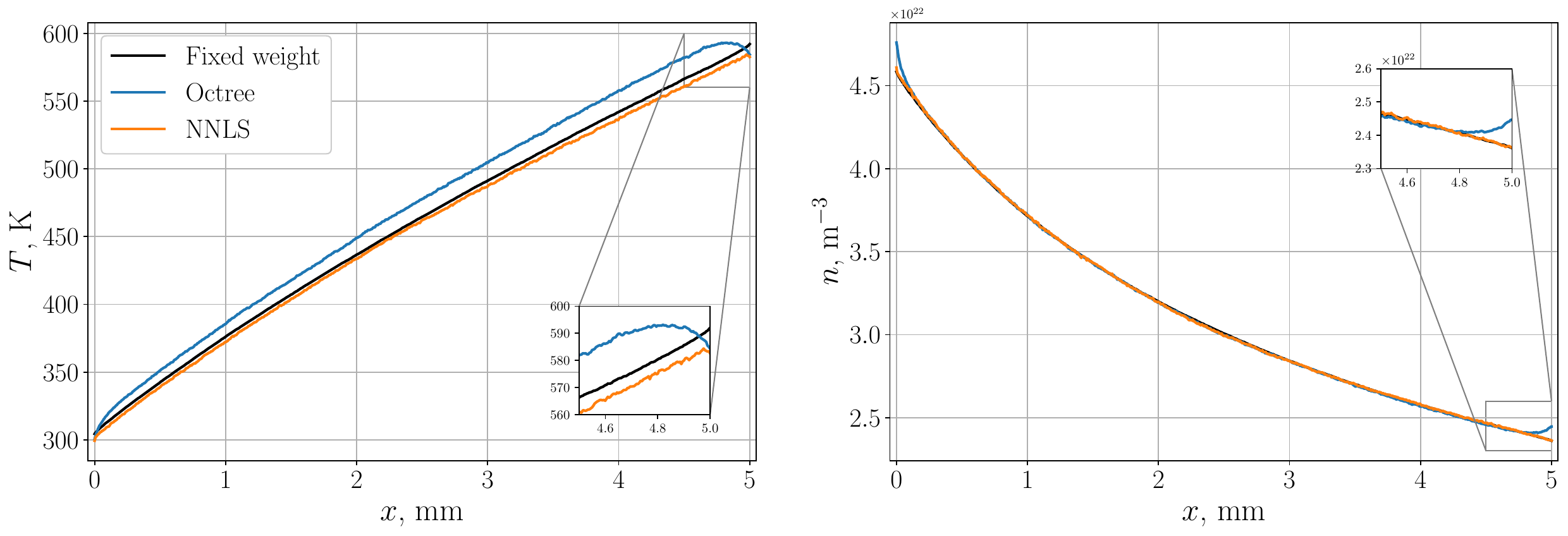}
  \caption{Time-averaged temperature (left) and number density (right) profiles.}\label{fig:fourier-T-ndens}
  \end{figure}

  Figure~(\ref{fig:fourier-T-ndens}) shows the time-averaged profiles of temperature and number density for the octree and NNLS merging approaches where the average
  number of particles per cell is equal to approximately 55. We see that the temperature results obtained with the NNLS approach
  lies much closer to the reference fixed-weight solution, whereas use of the octree-based merging leads to overestimation of the temperature in the whole flowfield and larger temperature gradients at the walls. For the number density in the bulk, both methods provide very
  similar results in good agreement with the reference solution, but near the walls, octree merging exhibits strong deviations from the reference solution.

\begin{figure}[t]
  \centering
  \includegraphics[width=0.98\textwidth]{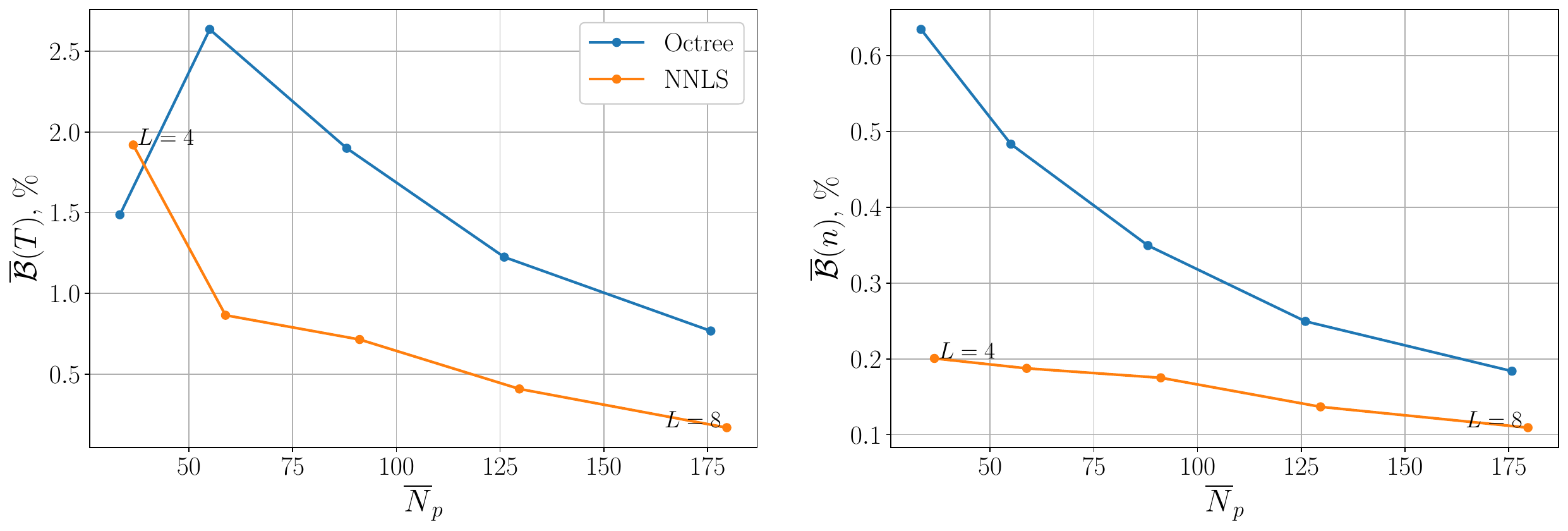}
  \caption{Relative bias of the time-averaged temperature (left) and number density (right) as a function
  of the average number of particles.}\label{fig:bias-T-ndends}
  \end{figure}

To compare approaches across various average numbers of particles, we consider the relative bias of temperature and number density with respect to the reference solutions, as shown on Figure~(\ref{fig:bias-T-ndends}). We see that the solutions obtained with the NNLS
merging exhibit lower bias throughout all particle counts, and present a noticeable improvement over those obtained via
use of the octree-based merging, especially at lower particle counts.

\begin{figure}[t]
  \centering
  \includegraphics[width=0.49\textwidth]{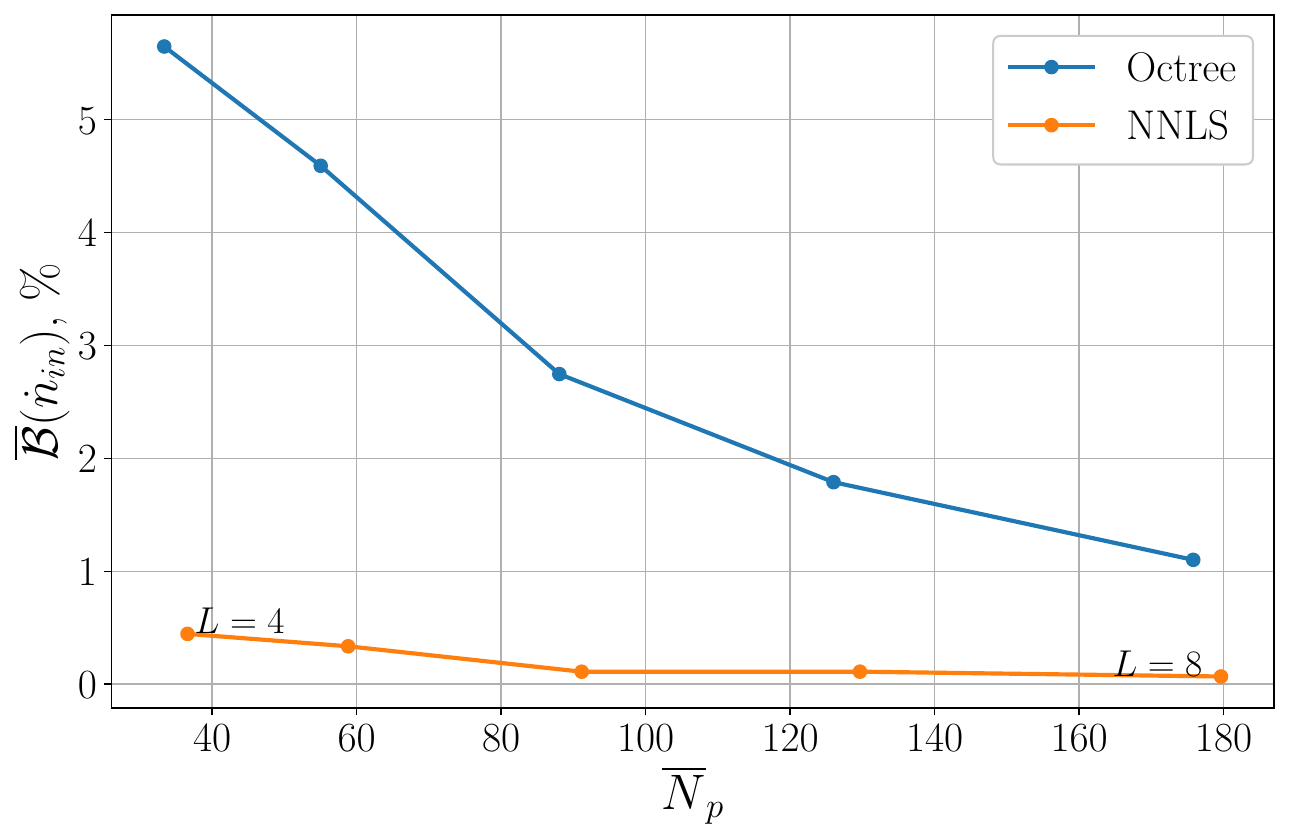}
  \caption{Relative bias of the incident number density flux at the right wall as a function
  of the average number of particles.}\label{fig:bias-incident-flux}
  \end{figure}

\begin{figure}[t]
  \centering
  \includegraphics[width=0.49\textwidth]{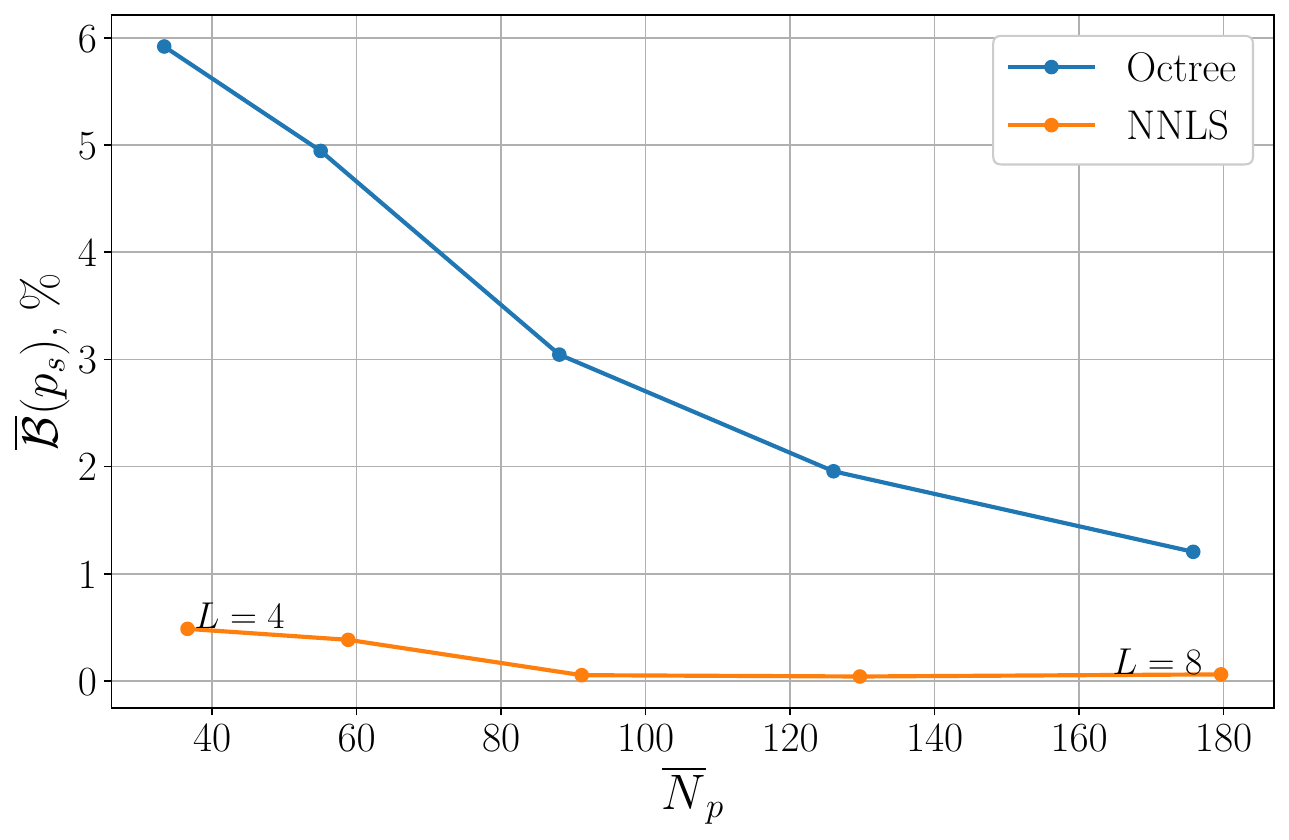}
  \caption{Relative bias of the pressure at the right wall as a function
  of the average number of particles.}\label{fig:bias-pressure}
  \end{figure}

\begin{figure}[t]
  \centering
  \includegraphics[width=0.49\textwidth]{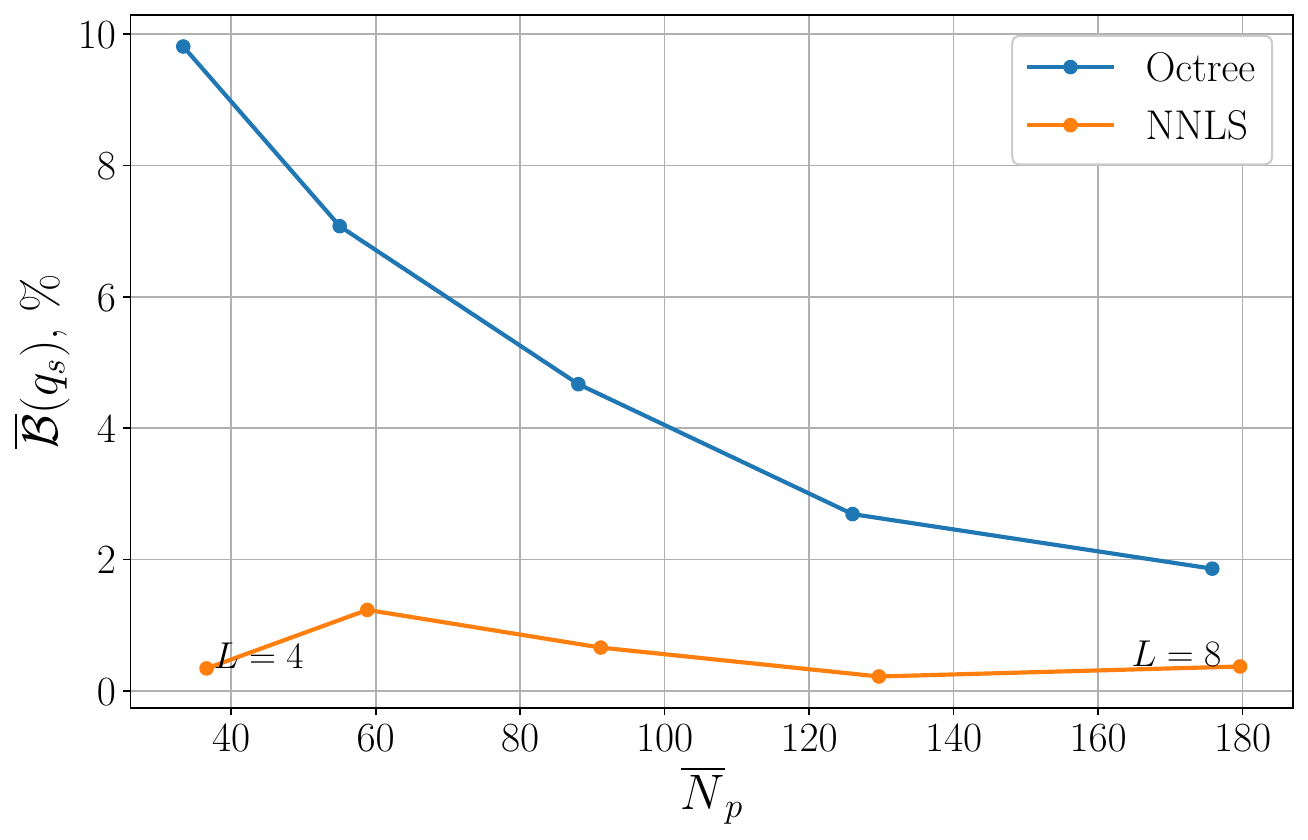}
  \caption{Relative bias of the kinetic energy flux at the right wall as a function
  of the average number of particles.}\label{fig:bias-heat-flux}
  \end{figure}
  Finally, we consider bias in the surface quantities, such as incident number density flux, pressure, and kinetic energy flux. Since the relative biases
  on the left and right walls were found to be very similar for all of the quantities, we only consider the bias at the right wall.

  Figures~\ref{fig:bias-incident-flux}--\ref{fig:bias-heat-flux} show the relative biases in the incident number density flux, surface pressure, and kinetic energy flux,
  respectively. We see that for all of the considered quantities, use of octree merging leads to significantly higher errors when compared to the NNLS-based merging results,
  especially at lower particle counts. This is especially pronounced for the higher-order moments (i.e. kinetic energy flux), where at low particle counts
  the time-averaged kinetic energy flux computed with use of octree merging is off by up to 10\%, whereas the error in the flux when using NNLS
  based merging ranges between 0.5\% and 1\%.

  Thus, we can conclude that use of NNLS merging in inhomogeneous simulations allows for use of lower average particle counts whilst still
  achieving lower bias error in key macroscopic quantities.

\subsection{Computational cost analysis}
\blue{Finally, we analyse the computational cost of the NNLS merging method and compare it to the octree merging method. We consider two of the
testcases presented above, namely the spatially homogeneous ionization and the one-dimensional Fourier flow. We analyze both the cost of the merging
procedure itself, as well as the cost of a full timestep (excluding computation of physical properties and  I/O),
i.e. collisions, merging, convection, particle sorting. All results were obtained on a workstation with a Intel Core i9-13900K processor and 128 GB of RAM.
}

\begin{figure}[t]
  \centering
  \includegraphics[width=0.98\textwidth]{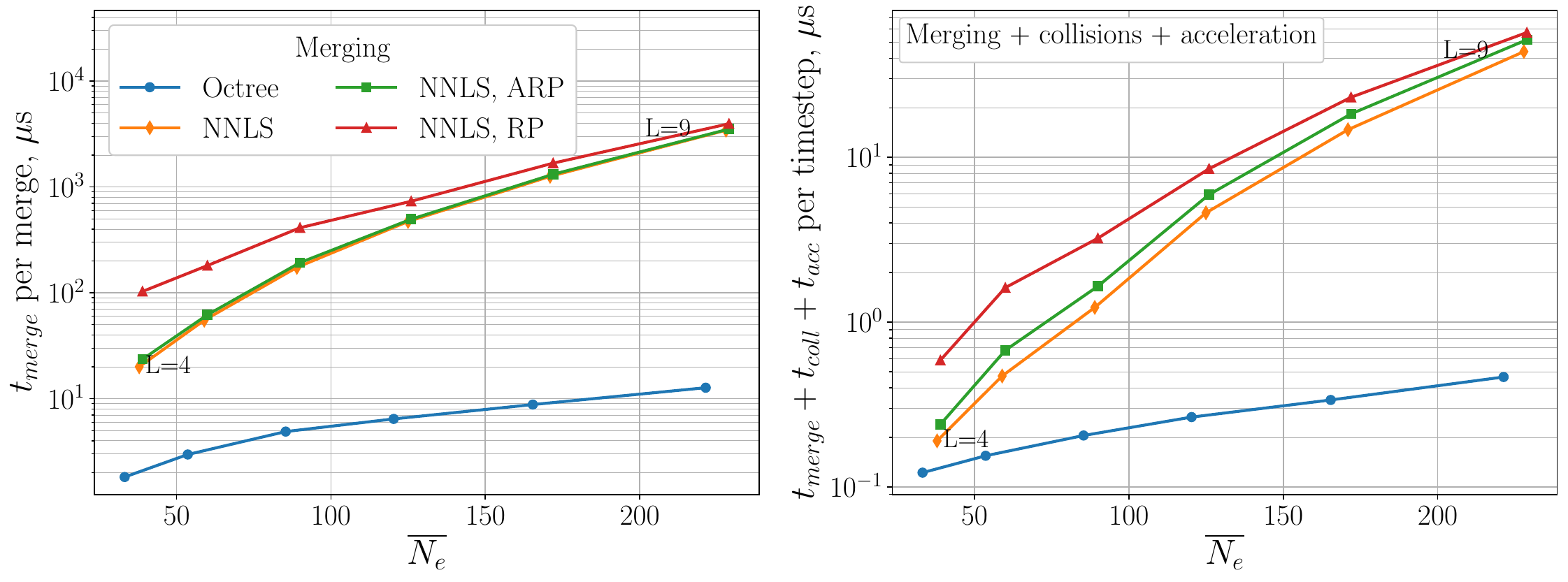}
  \caption{\blue{Cost of a single merge (left) and average cost of a single simulation timestep (right) as a function of the average
  number of particles in the simulation, spatially homogeneous ionization test case.}}\label{fig:cost-ionization}
  \end{figure}

\blue{First, we consider the spatially homogeneous ionization case discussed in~\ref{sec:ion}. Figure~\ref{fig:cost-ionization} shows the cost of a single merge and of a single simulation timestep (collisions, acceleration, and merging) for the ionization test case as a function of the average number of particles in the simulation. We see that although NNLS merging by itself is significantly more expensive than octree merging, at lower particle counts, the difference is not as pronounced when one considers other parts of the simulation. We also observe that exact preservation of merging rates carries a non-negligible cost, as evidenced by the ``NNLS, RP'' lying higher than the other NNLS lines.}

\begin{figure}[t]
  \centering
  \includegraphics[width=0.98\textwidth]{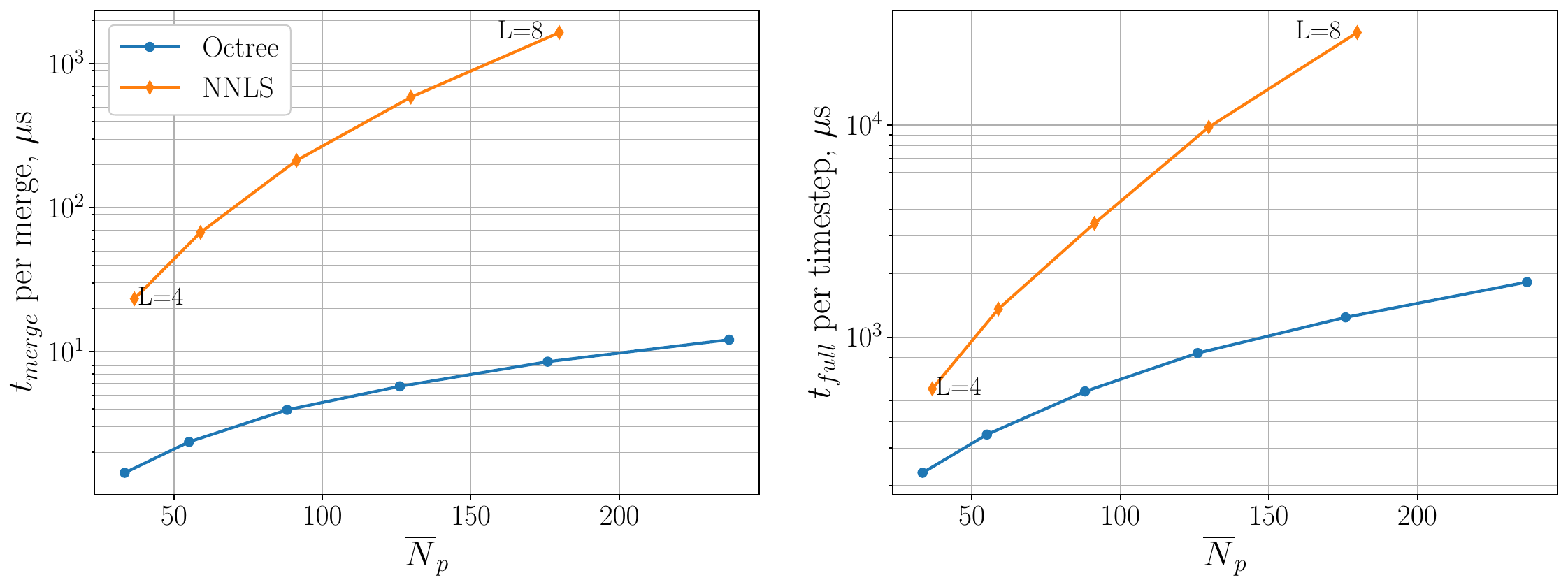}
  \caption{\blue{Cost of a single merge (left) and average cost of a single simulation timestep (right) as a function of the average
  number of particles in the simulation, Fourier flow test case.}}\label{fig:cost-fourier}
  \end{figure}

\blue{Figure~\ref{fig:cost-fourier} shows the cost of a single merge and of a single simulation timestep (collisions, acceleration, and merging) for the Fourier flow case detailed in~\ref{sec:fourier}. The cost of a single simulation timestep here includes collisions, particle convection, sorting, and merging. Similar to the ionization test case, the NNLS merging exhibits a noticeably higher computational cost than the octree merging, although
when the cost of other operations is included, the difference is less pronounced. However, we also note that cost as a function of the number of particles is not the only criterion for comparison, as the NNLS merging algorithm is able to achieve a better accuracy than the octree merging algorithm for a smaller number of particles. Therefore, we also plot the error in the wall pressure (previously shown on Figure~\ref{fig:bias-pressure} as a function of the average number of particles in the simulation) as a function of the average cost of performing
a single simulation timestep.}

\begin{figure}[t]
  \centering
  \includegraphics[width=0.49\textwidth]{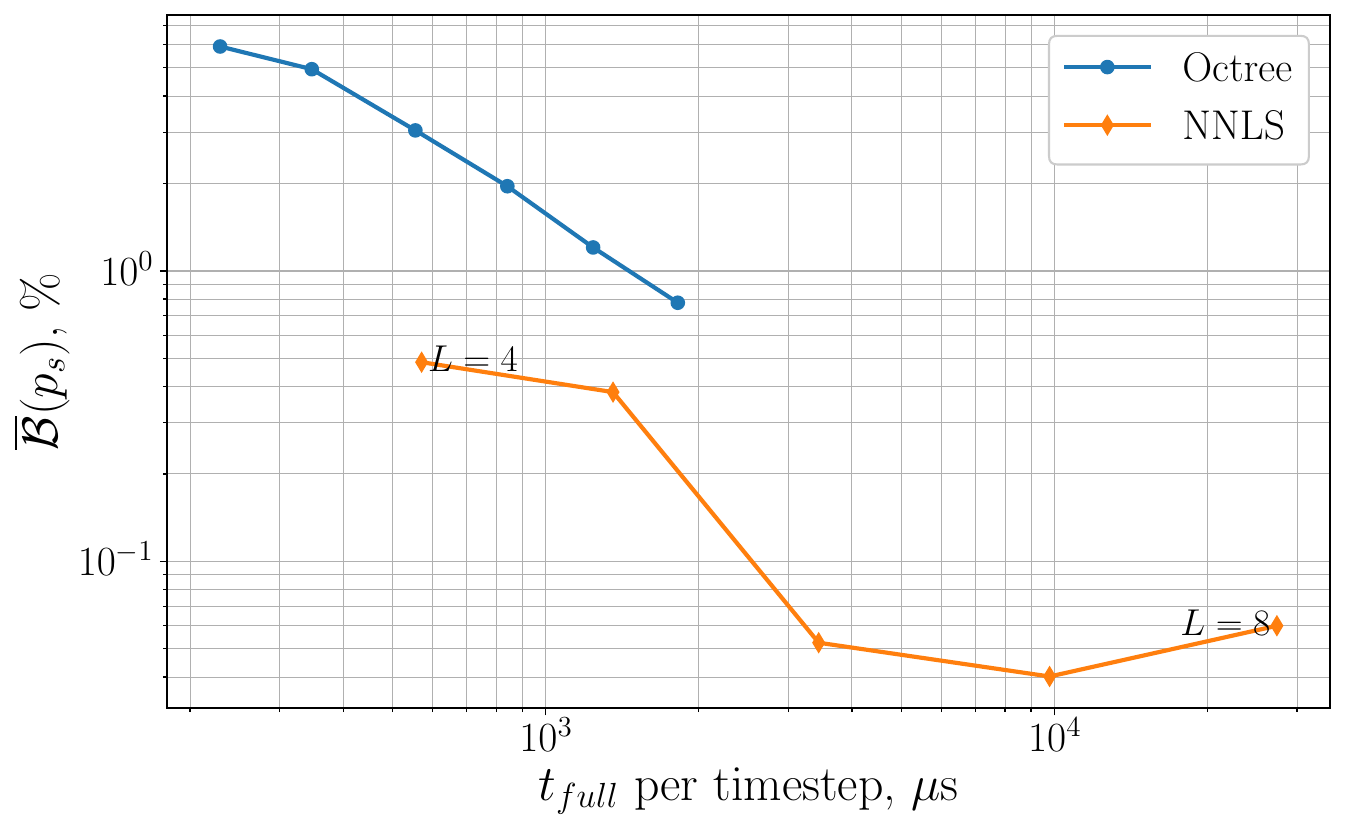}
  \caption{\blue{Error in the wall pressure as a function of the average
  cost a single simulation timestep, Fourier flow test case.}}\label{fig:error-vs-cost-fourier}
  \end{figure}

  \blue{Figure~\ref{fig:error-vs-cost-fourier} shows the error in the wall pressure as a function of the average cost of a single simulation timestep. We see that when considered from this viewpoint, NNLS-based merging outperforms the octree-based merging algorithm: even with relatively few particles per cell ($L=4$) and subsequently, a lower computational cost per timestep, NNLS-based merging achieves better accuracy than the octree-based merging with a higher particle count. We note that this metric, although it accounts for the memory consumption (since most of the computational cost is proportional to the number of particles used), it does so implicitly. In larger-scale simulations, being able to run at fewer particles per cell whilst achieving same or lower error, which is what the NNLS merging approach enables in this case, would be of additional benefit due to the cost of particle exchange between computational nodes when using parallelization.
  We therefore conclude that the proposed approach allows for more efficient simulations when viewed in term of errors-vs-computational cost, and requires fewer particles, thus reducing memory cost as well.}

\section{Conclusions}
We developed a new moment-preserving merging algorithm for homogeneous and inhomogeneous variable-weight DSMC simulations, that is based on solution
of a non-negative least squares problem. We extend the algorithm to conservation of exact and approximate inter-species collisional rates,
and apply it to several model rarefied gas dynamics problem, where it achieves better accuracy than the adaptive octree-based merging
algorithm.  
Possible extensions to the algorithm include its coupling with velocity space grouping approaches similar to the approach taken in~\cite{huerta2024situ}, extensions to particle-in-cell simulations, as well as the study of the potential speed-ups via use of other approaches to solving the NNLS problem~\cite{diakonikolas2022fast,chou2022non,bvelik2025efficient}.

\section{Acknowledgments}
The authors thank the German Research Foundation (DFG) for the financial support through 442047500/SFB1481 within the project B04 (Sparsity patterns in kinetic theory).



\end{document}